\documentclass[nocopyrightspace,nonatbib]{sigplanconf}

\pdfoutput=1
\newif\ifdraft\draftfalse
\ifdraft
  \usepackage[inline]{showlabels}
\fi

\usepackage{balance}

\usepackage{microtype}

\usepackage{setspace}

\usepackage{cmtt}

\usepackage{cite}

\usepackage{mathpartir}

\usepackage{mathdots}

\usepackage{stmaryrd}

\usepackage{textcomp}
\usepackage[clock]{ifsym}

\usepackage{marvosym}

\usepackage{booktabs}
\usepackage{dcolumn}
\newcolumntype{.}{D{.}{.}{-1}}

\usepackage{wrapfig}

\usepackage{nicefrac}

\def\clap#1{\hbox to 0pt{\hss#1\hss}}
\def\mathllap{\mathpalette\mathllapinternal}

\def\mathllapinternal#1#2{%
\llap{$\mathsurround=0pt#1{#2}$}}

\usepackage[flushleft,alwaysadjust]{paralist}

\usepackage{amsmath}
\usepackage{amssymb}

\newcommand\hcancel[2][black!60]{\setbox0=\hbox{$#2$}%
  \rlap{\raisebox{.35\ht0}{\textcolor{#1}{\rule{\wd0}{0.4pt}}}}#2}

\usepackage{dblfloatfix}

\usepackage[bottom,flushmargin]{footmisc}

\usepackage[font=bf,skip=2pt]{caption}
\usepackage[font=bf,captionskip=5pt]{subfig}
\makeatletter
\newbox\sf@box
\newenvironment{SubFloat}[2][]%
  {\def\sf@one{#1}%
   \def\sf@two{#2}%
   \setbox\sf@box\hbox
     \bgroup}%
  {  \egroup
   \ifx\@empty\sf@two\@empty\relax
     \def\sf@two{\@empty}
   \fi
   \ifx\@empty\sf@one\@empty\relax
     \subfloat[\sf@two]{\box\sf@box}%
   \else
     \subfloat[\sf@one][\sf@two]{\box\sf@box}%
   \fi}
\makeatother

\usepackage{url}

\usepackage{tikz}
\usetikzlibrary{decorations.pathreplacing}
\usetikzlibrary{positioning}
\pgfdeclarelayer{background}
\pgfdeclarelayer{foreground}
\pgfsetlayers{background,main,foreground}

\usepackage{delarray}

\definecolor{closure}{gray}{0.5}
\definecolor{tag}{gray}{0.8}

\renewcommand{\vdash}[2][8pt]{%
  \setbox0\vbox to #1{%
    {\color{#2}\leaders\vbox{\vskip1pt\hrule height 1pt width.5pt}\vfill}}%
  \ht0=6.5pt
  \box0
}

\newcolumntype{e}{c!{\vdash{closure}}}
\newenvironment{environ}[1]{%
  \ignorespaces%
  \begin{tabular}{#1}}{\end{tabular}\ignorespacesafterend}
\newcommand{\emptyenv}{\phantom{\vrule height6pt depth2pt width6pt}}
\newcommand{\closure}[2]{%
  \smash{%
  \begin{tikzpicture}[baseline=(env.base)]
      \node[inner sep=0pt,outer sep=0pt] (env) {#2};
      \begin{pgfonlayer}{background}
      \node[anchor=north east,fill=tag,font=\scriptsize,inner sep=0.4pt] (tag) at (env.north west) {#1};
          \draw[closure,rounded corners=1pt]
              (tag.south east)
              -- (tag.south west)
              -- (tag.north west)
              -- (tag.north east);
          \draw[closure,rounded corners=2pt]
              (env.north west)
              -- (env.north east)
              -- (env.south east)
              -- (env.south west)
              {[sharp corners] -- cycle};
      \end{pgfonlayer}
  \end{tikzpicture}}%
}

\usepackage{bigstrut}
\usepackage{array}
\usepackage{colortbl}
\usepackage{dcolumn}
\newcolumntype{.}{D{.}{.}{-1}}
\newcommand{\col}[1]{\ensuremath{\mathsf{#1}}}
\newcolumntype{H}{>{\columncolor{black}\color{white}}c}
\newcommand{\colhd}[1]{\multicolumn{1}{H}{\bigstrut\col{#1}}}
\newenvironment{littbl}{%
  \renewcommand{\arraystretch}{0.7}
  
  \ttfamily}{}
\newcommand{\tabname}[2]{%
  \multicolumn{#1}{@{}l}{%
    \begin{tikzpicture}[baseline]
      \node [inner sep=1pt] (T) {\phantom{#2}};
      \draw [fill=black]
            (T.south west) {[rounded corners=2pt] -- (T.north west)  --
            (T.north east)} -- (T.south east) -- cycle;
      \node [inner sep=1pt,text=white] {#2};
    \end{tikzpicture}}}

\usepackage{pmboxdraw}

\usepackage{nonfloat}

\makeatletter
\def\endarray{\crcr \egroup \egroup \@arrayright\gdef\@preamble{}\CT@end}
\makeatother

\DeclareRobustCommand\mathttfamily%
{\not@math@alphabet\ttfamily\mathtt%
\fontfamily\ttdefault\selectfont}

\usepackage{fancyvrb}
\usepackage{listings}
\makeatletter
\def\lst@visiblespace{\textnormal{\textvisiblespace\,}}
\makeatother
\lstMakeShortInline[basicstyle=\fontfamily\ttdefault\selectfont]|
\lstdefinelanguage{sql}{
  morecomment=[l]{--}
}
\lstdefinelanguage{xquery}{
  morecomment=[s]{(:}{:)}
}

\usepackage{refcount}

\usepackage{cleveref}

\crefname{section}{Section}{Sections}
\crefname{subsection}{Section}{Sections}
\crefname{subsubsection}{Section}{Sections}
\crefname{appendix}{Appendix}{Appendices}
\crefname{equation}{Equation}{Equations}
\crefname{figure}{Figure}{Figures}
\crefname{table}{Table}{Tables}
\crefname{subfigure}{Figure}{Figures}
\crefname{subtable}{Table}{Tables}
\crefformat{item}{\itshape #1.}
\newcounter{case}
\newcommand{\labelcase}[1]{\refstepcounter{case}\label[case]{#1}}
\creflabelformat{case}{(\oldstylenums{#1})}
\crefname{case}{case}{cases}

\newcommand{\args}[1]{\ensuremath{\smash[b]{\rlap{$_{#1}$}}}}

\newcommand{\meta}[1]{\textsf{#1}}

\newcommand{\Dispatch}[1]{\var{Dispatch}(#1)}

\DefineShortVerb{\|}
\SaveVerb{dollar}|$|\SaveVerb{dollar}|$|
\UndefineShortVerb{\|}
\newcommand{\vardollar}{\UseVerb{dollar}}

\newcommand{\XML}   {X\kern-0.07emM\kern-0.07emL}
\newcommand{\XPath} {X\kern-0.07emPath}
\newcommand{\XQuery}{X\kern-0.12emQue\-ry}
\newcommand{\XSLT}  {X\kern-0.07emS\kern-0.07emL\kern-0.07emT}
\newcommand{\SQL}   {S\kern-0.06emQ\kern-0.06emL}
\newcommand{\PLSQL} {P\kern-0.07emL\kern-0.03em/\kern-0.06emS\kern-0.06emQ\kern-0.06emL}
\newcommand{\Oracle}{Oracle~11\raisebox{3pt}{\textsl{g}}}
\newcommand{\TPCH}  {\mbox{TPC-H}}
\newcommand{\tpchcol}[2]{\sql{#1\string_#2}}

\newcommand{\zeroormore}[1]{#1^{\ast}}
\newcommand{\oneormore}[1]{#1^{\scriptscriptstyle+}}

\newcommand{\expr}{\ensuremath{\mathcal{E}}}

\newcommand{\query}{\ensuremath{\mathcal{Q}}}
\newcommand{\decl}{\ensuremath{\mathcal{D}}}
\newcommand{\repr}{\ensuremath{\mathcal{C}}}

\newcommand{\fun}[2]{#1\mathrel{\to}#2}

\newcommand{\clos}[1]{\tau_{#1}}

\newcommand{\xquery}[1]{\textmtt{#1}}
\newcommand{\sql}   [1]{\textmtt{#1}}
\newcommand{\var}   [1]{{\ensuremath{\mathit{#1}}}}

\begin{document}

\title{%
First-Class Functions for First-Order Database Engines}

\authorinfo
  {Torsten Grust\and Alexander Ulrich}
  {Universit\"at T\"ubingen, Germany}
  {\{torsten.grust,  alexander.ulrich\}@uni-tuebingen.de}

\maketitle
\thispagestyle{plain}

\begin{abstract}
  We describe \emph{query defunctionalization} which enables off-the-shelf
  first-order database engines to process queries over first-class
  functions. Support for first-class functions is cha\-racterized by
  the ability to treat functions like regular data items that can
  be constructed at query runtime, passed to or returned from other
  (higher-order) functions, assigned to variables, and stored in persistent
  data structures. Query defunctionalization is a non-invasive approach that
  transforms such function-centric queries into the data-centric operations
  implemented by common query processors. Experiments with \XQuery{} and
  \PLSQL{} database systems demonstrate that first-order database engines
  can faithfully and efficiently support the expressive
  ``\emph{functions as data}'' paradigm.
\end{abstract}

\section{Functions Should be First-Class}
\label{sec:first-class-functions}

Since the early working drafts of~2001, \XQuery's syntax and semantics
have followed a \emph{functional} style:\footnote{``[\dots] \XQuery{} is a
functional language in which a query is represented as an
expression.''~\cite[\S{}2]{InitialXQuery1.0}}
functions are applied to form complex expressions in a compositional fashion.
The resulting \XQuery{} script's top-level expression is
evaluated to return a sequence of items, \emph{i.e.}, atomic values or
\XML{} nodes~\cite{XQuery1.0}.

Ten years later, with the upcoming World Wide Web Consortium (W3C) \XQuery~3.0
Recommendation~\cite{XQuery3.0}, functions themselves now turn into
first-class items. Functions, built-in or user-defined, may be assigned
to variables, wrapped in sequences, or supplied as arguments to and
returned from higher-order functions. In effect, \XQuery{} finally
becomes a \emph{full-fledged functional language}.  Many useful
idioms are concisely expressed in this ``\emph{functions as data}'' paradigm.
We provide examples below and argue that support for first-class functions
benefits other database languages, \PLSQL{} in particular, as well.

This marks a significant change for query language implementations,
specifically for those built on top of (or right into) database kernels.
While atomic values, sequences, or \XML{} nodes are readily represented in
terms of the widespread first-order database data models~\cite{MonetDBXQuery},
this is less obvious for function items. Database kernels typically lack a
\emph{runtime} representation of functional values at all.
We address this challenge in the present work.

In query languages, the ``functions as data'' principle can surface in various
forms.

\medskip\noindent
\textbf{Functions as Values.}
\XQuery{}~3.0 introduces~$\var{name}\xquery{\#}n$ as notation to refer to the $n$-ary function
named~$\var{name}$: \xquery{math:pow\#2} refers to exponentiation while \xquery{fn:concat\#2}
denotes string concatenation, for example.  The values of these
expressions are functions---their types are of the
form~\xquery{function($t_{1}$) as $t_{2}$} or, more succinctly, $\fun{t_1}{t_2}$---which may be bound to
variables and applied to arguments. The evaluation of the expression
\xquery{let\,\$exp\,:=\,math:pow\#2\;return\;\$exp(2,3)} yields~\xquery{8},
for example.

\begin{figure}
  \centering\small
  \begin{lstlisting}[language=xquery]
declare function fold-right(
  $f as function(item(), item()*) as item()*,
  $z as item()*, $seq as item()*) as item()*
{
  if (empty($seq)) then $z else
    $f(fn:head($seq), fold-right($f, $z, fn:tail($seq)))
};
  \end{lstlisting}
  \caption{Higher-order function~\xquery{fold-right} (\XQuery~3.0).}
  \label{fig:xquery-fold-right}
\end{figure}

\medskip\noindent
\textbf{Higher-Order Functions.}
In their role of regular values, functions may be supplied as parameters to
and returned from other functions.  The latter, higher-order functions can
capture recurring patterns of computation and thus make for ideal building
blocks in query library designs.  Higher-order function~\xquery{fold-right}
is a prime example here---entire query language designs have been based on its
versatility~\cite{MapReduce,VersatileComprehensions}.
The \XQuery~3.0 variant~\xquery{fold-right(\$f,} \xquery{\$z,} \xquery{\$seq)}
is defined in~\cref{fig:xquery-fold-right}: it reduces a given input
sequence~$\xquery{\$seq} = \xquery{($e_{1}$,$e_{2}$,$\dots$,$e_{n}$)}$ to
the value \xquery{\$f($e_{1}$,\$f($e_{2}$,\$f($\dots$,\$f($e_{n}$,\$z)$\cdots$)))}.
Different choices for the functional parameter~\xquery{\$f} and~\xquery{\$z}
configure~\xquery{fold-right} to perform a variety of computations:
$$
\xquery{fold-right(math:pow\#2,\,1,\,($e_{1}$,$e_{2}$,$\dots$,$e_{n}$))}\qquad
$$%
(with numeric~$e_{i}$) computes the exponentiation tower
\smash[t]{$e_{1}^{\renewcommand{\arraystretch}{0.3}
                  \begin{array}{@{}c@{}c@{}c@{}}
                                       &                       & \scriptscriptstyle e_{n} \\[-5pt]
                                       &  \scriptscriptstyle \iddots  \\[-3pt]
                    \scriptstyle e_{2}
                  \end{array}}$},
while the expression
$$
\xquery{fold-right(fn:concat\#2, "", ($e_{1}$,$e_{2}$,$\dots$,$e_{n}$))}
$$%
will return the concatenation of the $n$ strings $e_{i}$.

\medskip\noindent
\textbf{Function Literals.}
Queries may use~\xquery{function($x$)$\,$\{$\,e\,$\}} to denote a literal
function (also: inline function or $\lambda$-expression $\lambda\kern0.05em x.e$).
Much like the literals of regular first-order types (numbers, strings, \dots),
function literals are pervasive if we adopt a functional mindset:

\begin{figure}[t]
  \centering\small
  \begin{lstlisting}[language=xquery]
declare function map:empty() {
  function($x) { () }                                     %\label{line:map:empty}%
};

declare function map:entry($k,$v) {
  function($map) {
     function($x) { if ($x = $k) then $v else $map($x) }  %\label{line:map:entry}%
  }
};

declare function map:new($es) {
  fold-right(function($f,$x) { $f($x) }, map:empty(), $es)
};

declare function map:remove($map,$k) {
    function($x) { if ($x = $k) then () else $map($x) }
};
  \end{lstlisting}
  \caption{Direct implementation of maps as functions from
    keys to values (\XQuery~3.0).}
  \label{fig:xquery-h-o-map}
\end{figure}

\begin{figure}[t]
  \centering\small
  \begin{lstlisting}[language=xquery]
(: wrap(), unwrap(): see %\lst@commentstyle\cref{sec:closure-representation-xquery}% :)

declare function map:empty() {
  element map {}
};

declare function map:entry($k, $v) {
  element entry {
    element key { wrap($k) }, element val { wrap($v) }
  }
};

declare function map:new($es) {
  element map { $es }
};

declare function map:get($map, $k) {
  unwrap($map/child::entry[child::key = $k][1]/
    child::val/child::node())
};

declare function map:remove($map, $k) {
  element map { $map/child::entry[child::key != $k] }
};
  \end{lstlisting}
  \caption{A first-order variant of \XQuery{} maps.}
  \label{fig:xquery-f-o-element-map}
\end{figure}

A map, or associative array, is a function from keys to values.
\cref{fig:xquery-h-o-map} takes this definition literally and implements
maps\footnote{Our design follows Michael Kay's proposal for maps
in \XSLT{}~3.0. Of two entries under
the same key, we return the entry inserted first (this is implementation-dependent:
\url{http://www.w3.org/TR/xslt-30/#map}).} in terms of functions.
Empty maps (created by~\xquery{map:empty}) are
functions that, for any key~\xquery{\$x}, will return the empty result~\xquery{()}.
A map with entry~\xquery{(\$k,\$v)} is a function that yields~\xquery{\$v} if a
key~$\xquery{\$x} = \xquery{\$k}$ is looked up (and otherwise will continue
to look for~\xquery{\$x} in the residual map~\xquery{\$map}).  Finally, \xquery{map:new(\$es)}
builds a complex map from a sequence of entries~\xquery{\$es}---an entry is added through
application to the residual map built so far.  As a consequence of this
implementation in terms of functions, lookups are idiomatically performed by applying
a map to a key, \emph{i.e.}, we may write
\noindent
    \begin{lstlisting}[language=xquery,numbers=none]
let $m := map:new((map:entry(1,"one"), map:entry(2,"two")))
return $m(2)  (: %\lst@commentstyle$\rightsquigarrow$% "two" :)
    \end{lstlisting}

An alternative, regular first-order implementation of maps is shown
in~\cref{fig:xquery-f-o-element-map}.  In this variant, map entries are
wrapped in pairs of~\xquery{key}/\xquery{val} \XML{} elements. A sequence of
such pairs under a common~\xquery{map} parent element forms a complex map.
Map lookup now requires an additional function~\xquery{map:get}---\emph{e.g.},
with~\xquery{\$m} as above: \xquery{map:get(\$m,2)}---that uses
\XPath{} path expressions to traverse the resulting \XML{} element hierarchy.
(We come back to~\xquery{wrap} and~\xquery{unwrap}
in~\cref{sec:closure-representation-xquery}.)

We claim that the functional variant in~\cref{fig:xquery-h-o-map} is not only
shorter but also clearer and arguably more declarative, as it represents  a
\emph{direct} realization of the ``a map is a function'' premise. Further,
once we study their implementation, we will see that the functional and
first-order variants ultimately lead the query processor to construct and
traverse similar data structures
(\cref{sec:closure-representation-xquery}). We gain clarity and elegance and
retain efficiency.

\begin{figure}
  \centering\small
  \begin{lstlisting}[language=sql,mathescape=true]
-- Based on (an excerpt of) the TPC-H schema:
--   ORDERS(o_orderkey, o_orderstatus, o_orderdate, %\textnormal{\dots}%)
--   LINEITEM(l_orderkey, l_shipdate, l_commitdate, %\textnormal{\dots}%)

-- determines the completion date of an order based on its items
CREATE FUNCTION item_dates(comp FUNCTION(DATE,DATE) RETURNS DATE)
    RETURNS (FUNCTION(ORDERS) RETURNS DATE) AS
BEGIN
 RETURN FUNCTION(o)                                           %\label{line:litfun3}%
        BEGIN RETURN (SELECT comp(MAX(li.l_commitdate),       %\label{line:applycomp}%
                                  MAX(li.l_shipdate))
                     FROM   LINEITEM li
                     WHERE  li.l_orderkey = o.o_orderkey);
   END;
END;

-- find completion date of an order based on its status
CREATE TABLE COMPLETION (                                %\label{line:tabstart}%
 c_orderstatus CHAR(1),
 c_completion  FUNCTION(ORDERS) RETURNS DATE);

INSERT INTO COMPLETION VALUES
 ('F', FUNCTION(o) BEGIN RETURN o.o_orderdate; END),     %\label{line:litfun1}%
 ('P', FUNCTION(o) BEGIN RETURN NULL; END),              %\label{line:litfun2}%
 ('O', item_dates(GREATEST));                            %\label{line:currying}\label{line:namegreatest}\label{line:tabend}%

-- determine the completion date of all orders
SELECT o.o_orderkey,                                     %\label{line:sqlstart}%
       o.o_orderstatus,
       c.c_completion(o) AS completion                   %\label{line:applycompletion}%
FROM   ORDERS o, COMPLETION c
WHERE  o.o_orderstatus = c.c_orderstatus;                %\label{line:sqlend}%
  \end{lstlisting}
  \caption{Using first-class functions in \PLSQL.}
  \label{fig:sql-h-o-PLSQL}
\end{figure}

\medskip\noindent \textbf{Functions in Data Structures.}
Widely adopted database programming languages,
notably~\PLSQL{}~\cite{OraclePLSQL}, treat functions as second-class citizens:
in particular, regular values may be stored in table cells while functions may
not.  This precludes a programming style in which queries combine tables of
functions and values in a concise and natural fashion.

The code of \cref{fig:sql-h-o-PLSQL} is written in a hypothetical dialect of
\PLSQL{} in which this restriction has been lifted.  In this dia\-lect,
the function type~$\fun{t_{1}}{t_{2}}$ reads
\sql{FUNCTION($t_{1}$)\;RETURNS\;$t_{2}$} and
\sql{FUNCTION($x$)\;BEGIN\,$e$\;END} denotes a literal function with
argument~$x$ and body~$e$.\footnote{We are not keen to propose
syntax here. Any notation that promotes first-class functions would be fine.}

The example code augments a \TPCH{} database~\cite{TPC-H} with a configurable
method to determine order completion dates.
In~\crefrange{line:tabstart}{line:tabend}, table~\sql{COMPLETION} is created
and populated with one possible configuration that maps an order status
(column~\tpchcol{c}{orderstatus}) to its particular method of completion
date computation.  These methods are specified as functions of type
~\sql{FUNCTION(ORDERS)\;RETURNS\;DATE}\footnote{Type~\sql{ORDERS} denotes
the type of the records in table~\sql{ORDERS}.} held in~\tpchcol{c}{completion},
a functional column:
while we directly return its~\tpchcol{o}{orderdate} value for a finalized order
(status~\sql{'F'}) and respond with an undefined
\sql{NULL} date for orders in processing~(\sql{'P'}), the completion date of an
open order~(\sql{'O'}) is determined by
function~\sql{\tpchcol{item}{dates}(GREATEST)}: this function consults the
commitment and shipment dates of the order's items and then returns the most
recent of the two (since argument~\sql{comp} is~\sql{GREATEST}).\footnote{Built-in
\SQL{} function~\sql{GREATEST} (\sql{LEAST}) returns the larger (smaller) of
its two arguments.}

Function~\tpchcol{item}{dates} itself has been designed to be
configurable. Its higher-order type
$$
\fun{(\fun{\sql{DATE} \times \sql{DATE}}{\sql{DATE}})}{(\fun{\sql{ORDERS}}{\sql{DATE}})}
$$
indicates that~\tpchcol{item}{dates} returns a function to calculate order
completion dates once it has been supplied with a suitable date comparator
(\emph{e.g.}, \sql{GREATEST} in~\cref{line:currying}). This
makes~\tpchcol{item}{dates} a \emph{curried} function which consumes its
arguments successively (date comparator first, order second)---a prevalent
idiom in function-centric programming~\cite{IntroFunctionalProgramming}.

Note that the built-in and user-defined functions~\sql{GREATEST}
and~\tpchcol{item}{dates} are considered values as are the two literal
functions in~\cref{line:litfun1,line:litfun2}.  As such they may be stored in
table cells---\emph{e.g.}, in column~\tpchcol{c}{completion} of
table~\sql{COMPLETION}---and then accessed by \SQL{} queries.  The query
in~\crefrange{line:sqlstart}{line:sqlend} exercises the latter and calculates
the completion dates for all orders based on the current
configuration~in~\sql{COMPLETION}.

Once more we obtain a natural solution in terms of first-class
functions---this time in the role of values that populate tables.  Queries
can then be used to combine functions and their arguments in flexible ways.
We have demonstrated further use cases for \PLSQL{} defunctionalization (including
offbeat examples, \emph{e.g.}, the simulation of algebraic data types)
in~\cite{FunctionsAreDataToo}.

\medskip\noindent
\textbf{Contributions.}  The present work shows that off-the-shelf
database systems can faithfully and efficiently support expressive query
languages that promote first-class functions.   Our specific contributions are
these:
\begin{compactitem}
  \item We apply \emph{defunctionalization} to queries, a source
    transformation that trades functional values for first-order values which
    existing query engines can process efficiently.
  \item We discuss representations of closures that fit database data models
    and take size and sharing issues into account.
  \item We demonstrate how these techniques apply to widely adop\-ted query
    languages (\XQuery, \PLSQL) and established systems (\emph{e.g.}, Oracle
    and Postgre\SQL).
  \item We show that defunctionalization introduces a tolerable runtime
    overhead (first-order queries are not affected at all) and how simple
    optimizations further reduce the costs.
\end{compactitem}
Defunctionalization is an
established technique in programming languages and it deserves to be
better known in the database systems arena.

\medskip\noindent
The approach revolves around the concept of \emph{closure} which we discuss
briefly in~\cref{sec:closures}.
\cref{sec:defunctionalization} shows how defunctionalization maps queries over
first-class functions to regular first-order constructs. We focus on \XQuery{}
first and then carry over to \PLSQL{} in~\cref{sec:defunctionalization-plsql}.
Issues of efficient closure representation are addressed
in~\cref{sec:closure-representation}.
\cref{sec:experiments} assesses the space and time overhead of defunctionalization
and discusses how costs may be kept in check. \cref{sec:related-work} reviews
related efforts before we conclude in~\cref{sec:conclusions}.

\section{Functions as Values: Closures}
\label{sec:closures}

This work deliberately pursues a non-invasive approach that enables
off-the-shelf database systems to support the function-centric style of queries
we have advocated in~\cref{sec:first-class-functions}.  If these existing
first-order query engines are to be used for evaluation, it follows that we require a
\emph{first-order representation} of functional values.
\emph{Closures}~\cite{ClosureOriginal,ClosureCodePointer} provide such a
representation. We very briefly recall the concept here.

\begin{figure}
  \centering\small
  \begin{lstlisting}[language=xquery]
declare function group-by($seq as item()*,
                          $key as function(item()*) as item()*)
  as (function() as item()*)*
{
  let $keys := for $x in $seq return $key($x) %\label{line:dynamic-call}%
  for $k in distinct-values($keys)
  return
    function() { $seq[$key(.) = $k] } %\label{line:capture}%
};

let $fib := (0,1,1,2,3,5,8,13,21,34) %\label{line:fib}%
for $g in group-by($fib, function($x) { $x mod 2 }) %\label{line:mod}%
return
  element group { $g() }  %\label{line:arg-less-call}%
  \end{lstlisting}
  \caption{A grouping function that represents the individual groups
    in terms of closures (\XQuery~3.0).}
  \label{fig:xquery-h-o-group-by}
\end{figure}

The \XQuery{}~3.0 snippet of \cref{fig:xquery-h-o-group-by} defines the higher-order
grouping function~\xquery{group-by} which receives the grouping criterion in
terms of the functional argument~\xquery{\$key}: a group is the sequence of
those items~\xquery{\$x} in~\xquery{\$seq} that map to the same key value~\xquery{\$key(\$x)}.
Since \XQuery{} implicitly flattens nested sequences, \xquery{group-by} cannot
directly yield the sequence of all groups.  Instead, \xquery{group-by} returns
a sequence of functions each of which, when applied to zero arguments, produces
``its'' group.  The sample code in~\crefrange{line:fib}{line:arg-less-call}
uses~\xquery{group-by} to partition the first few elements of the Fibonacci series into
odd/even numbers and then wraps the two resulting groups in \XML{} \xquery{group} elements.

\medskip\noindent
\textbf{Closures.} Note that the inline function definition in
line~\cref{line:capture} captures the values of the free variables~\xquery{\$k},
\xquery{\$key}, and~\xquery{\$seq} which is just the information required to
produce the group for key~\xquery{\$k}.  More general, the language implementation
will represent a functional value~$f$ as a bundle that comprises
\begin{compactenum}[(1)]
\item the \emph{code} of $f$'s body and
\item its \emph{environment}, \emph{i.e.}, the bindings of the body's free
  variables at the time $f$ was defined.
\end{compactenum}
Together, code and environment define the \emph{closure} for function~$f$.
In the sequel, we will use
$$
\closure{$\ell$}{
  \begin{environ}{eec}
    $x_{1}$ & $\cdots$ & $x_{n}$
  \end{environ}
}
$$%
to denote a closure whose environment contains $n \geqslant 0$~free variables
$v_{1},\dots,v_{n}$ bound to the values~$x_{1},\dots,x_{n}$.\footnote{If we agree on a variable
order, there is no need to save the variable names~$v_{i}$
in the environment.} Label~$\ell$ identifies the code of the
function's body (in the original work on closures, code pointers
were used instead~\cite{ClosureCodePointer}). In the example
of~\cref{fig:xquery-h-o-group-by}, two closures are constructed
at~\cref{line:capture} (there are two distinct grouping
keys~$\xquery{\$k} = \xquery{0}, \xquery{1}$) that represent instances
of the literal function. If we order the free variables as~\xquery{\$k},
\xquery{\$key}, \xquery{\$seq}, these closures read
$$
\closure{$\ell_{1}$}{
  \begin{environ}{eec}
    \xquery{0} & \closure{$\ell_{2}$}{\emptyenv} & \xquery{(0,1,1,2,$\dots$)}
  \end{environ}
}
\quad\text{and}\quad
\closure{$\ell_{1}$}{
  \begin{environ}{eec}
    \xquery{1} & \closure{$\ell_{2}$}{\emptyenv} & \xquery{(0,1,1,2,$\dots$)}
  \end{environ}
}
\enskip.
$$%
(the two closures share label~$\ell_{1}$ since both refer to the
  same body code~\xquery{\$seq[\$key(.)\,=\,\$k]}). Observe that
\begin{itemize}
\item closures may be \emph{nested}: \xquery{\$key} is bound to
  closure~\closure{$\ell_{2}$}{\emptyenv} with empty environment, representing the
  literal~\xquery{function(\$x)\,\{\,\$x\,mod\,2\,\}} (defined in~\cref{line:mod})
  whose body has no free variables, and
\item closures may \emph{contain and share data of significant size}: both closures
  contain a copy of the~\xquery{\$fib} sequence (since free variable~\xquery{\$seq}
  was bound to~\xquery{\$fib}).
\end{itemize}
We will address issues of closure nesting, sharing, and size
in~\cref{sec:closure-representation,sec:experiments}.

\medskip\noindent
The key idea of \emph{defunctionalization}, described next, is to
trade functional values for their closure representation---ultimately, this
leaves us with an equivalent first-order query.

\section{Query Defunctionalization}
\label{sec:defunctionalization}

\emph{Query defunctionalization} is a source-level transformation that
translates queries over first-class functions into equivalent first-order
queries.  Here, our discussion revolves around \XQuery{} but
defunctionalization is readily adapted to other query languages, \emph{e.g.},
\PLSQL{} (see~\cref{sec:defunctionalization-plsql}).

The source language is \XQuery~3.0, restricted to the constructs that
are admitted by the grammar of~\cref{fig:source-grammar} (these restrictions
aid brevity---defunctionalization is straightforwardly extended to cover the
full \XQuery~3.0 specification). Notably, the language subset includes
\begin{compactitem}
\item two kinds of expressions that yield \emph{functional values} (literal
  functions of the form \xquery{function(\$$x_{1}$,$\dots$,\$$x_{n}$)\,\{\,$e$\,\}}
  as well as named function references~\xquery{$\var{name}$\#$n$}), and
\item \emph{dynamic function calls} of the
  form~\xquery{$e$($e_{1}$,$\dots$,$e_{n}$)}, in which expression~$e$
  evaluates to an $n$-ary function that is subsequently applied to the
  appropriate number of arguments.
\end{compactitem}

\begin{figure}
  \centering\small
  $
  \begin{array}{@{}rcl}
    \var{Program} & \to &
    \var{\zeroormore{\var{FunDecl}}}~\var{Expr}
    \\
    \var{FunDecl} & \to &
    |declare|~|function|~\var{QName}|(|\zeroormore{\vardollar{}\!\var{Var}}|)|~|{|~\var{Expr}~|};|
    \\
    \var{Expr} & \to &
    |for|~\vardollar{}\!\var{Var}~|in|~\var{Expr}~|return|~\var{Expr}
    \\
    & \mid &
    |let|~\vardollar{}\!\var{Var}~|:=|~\var{Expr}~|return|~\var{Expr}
    \\
    & \mid &
    \vardollar{}\!\var{Var}
    \\
    & \mid &
    |if|~|(|\var{Expr}|)|~|then|~\var{Expr}~|else|~\var{Expr}
    \\
    & \mid &
    |(| \zeroormore{Expr} |)|
    \\
    & \mid & \var{Expr} |/| \var{Axis} |::| \var{NodeTest}
    \\
    & \mid &
    |element|~\var{QName}~|{|~Expr~|}|
    \\
    & \mid &
    \var{Expr}|[|\var{Expr}|]|
    \\
    & \mid &
    |.|
    \\
    & \mid &
    \var{QName}\,|(|\zeroormore{\var{Expr}}|)|
    \\
    & \mid &
    |function|~|(|\zeroormore{\vardollar{}\!\var{Var}}|)|~|{|~\var{Expr}~|}|
    \\
    & \mid &
    \var{QName}|#|\var{IntegerLiteral}
    \\
    & \mid &
    \var{Expr}\,|(|\zeroormore{\var{Expr}}|)|
    \\
    & \mid &
    \cdots
    \\
    \var{Var} & \to &
    \var{QName}
  \end{array}
  $
  \caption{%
    Relevant \XQuery{} subset (source language), excerpt of the XQuery~3.0
    Candidate Recommendation~\protect\cite{XQuery3.0}.}
\label{fig:source-grammar}
\end{figure}

\begin{figure}
  \centering\small
  $
  \renewcommand{\arraystretch}{1.2}%
  \begin{array}{rcl}
   \var{Expr} & \to &
   [~\emph{constructs of~\cref{fig:source-grammar}}~]
   \\
   & \mid &
   \hcancel{\texttt{function~($\zeroormore{\vardollar{}\!\var{Var}}$)~\{~\var{Expr}~\}}}
   \\
   & \mid &
   \hcancel{\texttt{\var{QName}\#\var{IntegerLiteral}}}
   \\
   & \mid &
   \hcancel{\texttt{\var{Expr}\,($\zeroormore{\var{Expr}}$)}}
   \\
   & \mid &
   \closure{$\ell_{}$}{
     \begin{environ}{eec}
       \var{Expr} & $\cdots$ & \var{Expr}
     \end{environ}
   }
   \\
   & \mid &
   \meta{case}~\var{Expr}~\meta{of}~\oneormore{\var{Case}}
   \\[2ex]
   \var{Case}
   & \to &
   \closure{$\ell_{}$}{
     \begin{environ}{eec}
       \vardollar\!\var{Var} & $\cdots$ & \vardollar\!\var{Var}
     \end{environ}
   }~\Rightarrow~\var{Expr}
  \end{array}
  $
\caption{%
  Target language: functional values and dynamic function calls
  are removed. New: closure construction and elimination.}
\label{fig:grammar-plus}
\end{figure}

\medskip\noindent
The transformation target is a first-order dialect of~\XQuery{}~1.0 to which
we add closure construction and elimination.
A \emph{closure constructor}
\closure{$\ell$}{
  \begin{environ}{eec}
    $x_{1}$ & $\cdots$ & $x_{n}$
  \end{environ}
}
builds a closure with label~$\ell$ and an environment of values~$x_{1}, \dots, x_{n}$.
\emph{Closure elimination}, expressed using~$\meta{case} \cdots \meta{of}$, discriminates
on a closure's label and then extracts the environment contents: from
the $b$~branches in the expression%
\[
\begin{array}{@{\mskip25mu}r@{~}c@{~}l}
  \multicolumn{1}{l}{\meta{case}\,e\,\meta{of}}
  \\
  \closure{$\ell_{1}$}{
    \begin{environ}{eec}
      \xquery{\$$v_{1,1}$} & $\cdots$ & \xquery{\$$v_{1,n_{1}}$}
    \end{environ}
  } & \Rightarrow & e_{1}
  \\[-2pt]
  \multicolumn{1}{@{\mskip50mu}l}{\vdots}
  \\[0pt]
  \closure{$\ell_{b}$}{
    \begin{environ}{eec}
      \xquery{\$$v_{b,1}$} & $\cdots$ & \xquery{\$$v_{b,n_{b}}$}
    \end{environ}
  } & \Rightarrow & e_{b} \enskip,\\[2ex]
\end{array}
\]%
if $e$ evaluates to the closure
\closure{$\ell_{i}$}{
\begin{environ}{eec}
    $x_{1}$ & $\cdots$ & $x_{n}$
\end{environ}},
$\meta{case}\cdots\meta{of}$ will pick the $i\mskip1mu$th branch and evaluate $e_{i}$
with the variables~\xquery{\$$v_{i,j}$} bound to the values~$x_{j}$.
We discuss ways to express the construction and elimination of closures in terms
of regular query language constructs in \cref{sec:closure-representation}.

\cref{fig:grammar-plus} shows the relevant excerpt of the resulting
target language. In a sense, this modified grammar captures the essence of
defunctionalization: functional values and dynamic function calls are
traded for the explicit construction and elimination of first-order
closures.  The translation can be sketched as follows:
\begin{compactenum}
\item[\labelcref{case:literal}] A literal function is replaced by a closure constructor whose
  environment is populated with the bindings of the free variables referenced
  in the function's body.  The body's code is wrapped inside a new top-level
  surrogate function~$\ell$ whose name also serves as the closure label.
  \label{item:case1}
\item[\labelcref{case:name}] A reference to a function named~$\ell$  is replaced by a closure constructor
  with empty environment and label~$\ell$.
\item[\labelcref{case:apply}] A dynamic function call (now equivalent to an application of a closure
  with label~$\ell$ to zero or more arguments) is translated into a static
  function call to a generated \emph{dispatcher} function. The dispatcher receives
  the closure as well as the arguments and then uses closure elimination to forward
  the call to function~$\ell$, passing the environment contents (if any) along with
  the arguments.
\end{compactenum}

\medskip\noindent
\cref{sec:defunctionalization-for-xquery} elaborates the details of this
transformation, including the generation of dispatchers, for the \XQuery{}
case.  A syntax-directed top-down traversal identifies the relevant spots in a
given program at which closure introduction or elimination has to be performed
according to the cases~\labelcref{case:literal} to~\labelcref{case:apply}
above.  All other program constructs remain unchanged. The application of
defunctionalization to the \XQuery{} program of~\cref{fig:xquery-h-o-group-by}
yields the code of~\cref{fig:xquery-f-o-group-by}.  We find the expected
surrogate functions~$\ell_{1,2}$, dispatchers (\xquery{dispatch\_$n$}), and
static dispatcher invocations. Overall, the resulting defunctionalized query
adheres to the target language of~\cref{fig:grammar-plus}, \emph{i.e.}, the
query is first-order. Once we choose a specific implementation for closure
construction and elimination, we obtain a query that may be executed by any
\XQuery{}~1.0 processor.

\begin{figure}
  \centering\small
  \begin{lstlisting}[language=xquery]
declare function %$\ell_{2}$%($x) { $x mod 2 }; %\label{line:surrogate-start}%
declare function %$\ell_{1}$%($k, $key, $seq) {
  $seq[(dispatch_1($key, .)) = $k]
};                                              %\label{line:surrogate-end}%
declare function dispatch_0($clos) {
  %\meta{case}% $clos %\meta{of}%
    %$\closure{$\ell_{1}$}{
        \begin{environ}{eec}
          \$k & \$key & \$seq
        \end{environ}} \Rightarrow$% %$\ell_{1}$%($k, $key, $seq)
};
declare function dispatch_1($clos, $b1) {
  %\meta{case}% $clos %\meta{of}%
    %$\closure{$\ell_{2}$}{\emptyenv} \Rightarrow$% %$\ell_{2}$%($b1)
};
declare function group-by($seq, $key) { %\label{line:group-by-f-o-group-by-start}%
  let $keys := for $x in $seq return dispatch_1($key, $x) %\label{line:group-by-dispatch-1}%
  for $k in distinct-values($keys)
  return %$\closure{$\ell_{1}$}{
             \begin{environ}{eec}
               \$k & \$key & \$seq
             \end{environ}
         }$% %\label{line:capture-closure}%
}; %\label{line:group-by-f-o-group-by-end}%

let $fib := (0,1,1,2,3,5,8,13,21,34)
for $g in group-by($fib, %$\closure{$\ell_{2}$}{\emptyenv}$%) %\label{line:mod-closure}%
return
  element group { dispatch_0($g) }
  \end{lstlisting}
  \caption{Defunctionalized first-order variant of the \XQuery{}~\xquery{group-by}
    example in~\cref{fig:xquery-h-o-group-by}.}
  \label{fig:xquery-f-o-group-by}
\end{figure}


\subsection{Query Defunctionalization for \PLSQL{}}
\label{sec:defunctionalization-plsql}

Query defunctionalization does not need to be reinvented if we carry
it over to \PLSQL{}. Much like for \XQuery{}, the defunctionalization
transformation for a
\PLSQL{} dialect with first-class functions builds on three core
cases (see above and~\cref{fig:defun-rewrite-comprehension}
in~\cref{sec:defunctionalization-for-xquery}):
\begin{compactenum}
\item[\labelcref{case:literal}] the creation of function literals
  (applies in~\cref{line:litfun1,line:litfun2,line:litfun3} of the
  \PLSQL{} example in~\cref{fig:sql-h-o-PLSQL}),
\item[\labelcref{case:name}] references to named function
  values (\sql{GREATEST} in~\cref{line:namegreatest}), and
\item[\labelcref{case:apply}] dynamic function application
  (applies in~\cref{line:applycomp,line:applycompletion}).
\end{compactenum}

\begin{figure}[t]
  \centering\small
  \begin{lstlisting}[language=sql,mathescape=true]
CREATE FUNCTION %$\ell_{1}$%(o ORDERS, comp %$\clos{\fun{\sql{DATE} \times \sql{DATE}}{\sql{DATE}}}$%) RETURNS DATE AS%\label{line:l1}%
BEGIN
  RETURN (SELECT dispatch_2(comp, MAX(li.l_commitdate),%\label{line:innerSFW}%
                                  MAX(li.l_shipdate))
          FROM   LINEITEM li
          WHERE  li.l_orderkey = o.o_orderkey);
END;

CREATE FUNCTION %$\ell_{2}$%(o ORDERS) RETURNS DATE AS%\label{line:l2}%
BEGIN
  RETURN o.o_orderdate;
END;

CREATE FUNCTION %$\ell_{3}$%(o ORDERS) RETURNS DATE AS%\label{line:l3}%
BEGIN
  RETURN NULL;
END;

CREATE FUNCTION dispatch_1(clos %$\clos{\fun{\sql{ORDERS}}{\sql{DATE}}}$%, b1 ORDERS)%\label{line:dispatch1}%
    RETURNS DATE AS
BEGIN
  %\meta{case}% clos %\meta{of}%
    %\vrule depth2.5pt width0pt%%$\closure{$\ell_{1}$}{\begin{environ}{c}comp\end{environ}}\mskip-1mu$% %$\Rightarrow$% %$\ell_{1}$%(b1, comp)
    %\vrule depth2.5pt width0pt%%$\closure{$\ell_{2}$}{\emptyenv}$%     %$\Rightarrow$% %$\ell_{2}$%(b1)
    %$\closure{$\ell_{3}$}{\emptyenv}$%     %$\Rightarrow$% %$\ell_{3}$%(b1)
END;

CREATE FUNCTION dispatch_2(clos %$\clos{\fun{\sql{DATE} \times \sql{DATE}}{\sql{DATE}}}$%, b1 DATE, b2 DATE)%\label{line:dispatch2}%
    RETURNS DATE AS
BEGIN
  %\meta{case}% clos %\meta{of}%
    %$\closure{$\ell_{4}$}{\emptyenv} \Rightarrow$% GREATEST(d1, d2)
END;

CREATE FUNCTION item_dates(comp %$\clos{\fun{\sql{DATE} \times \sql{DATE}}{\sql{DATE}}}$%) %\label{line:itemdates}%
    RETURNS %$\clos{\fun{\sql{ORDERS}}{\sql{DATE}}}$% AS
BEGIN
  RETURN %$\closure{$\ell_{1}$}{\begin{environ}{c}comp\end{environ}}$%;
END;

CREATE TABLE COMPLETION (
  c_orderstatus CHAR(1),
  c_completion  %$\clos{\fun{\sql{ORDERS}}{\sql{DATE}}}$%);

INSERT INTO COMPLETION VALUES                                                   %\label{line:popstart}%
  %\vrule depth2.5pt width0pt%('F', %\closure{$\ell_{2}$}{\emptyenv}%),
  %\vrule depth2.5pt width0pt%('P', %\closure{$\ell_{3}$}{\emptyenv}%);
  ('O', item_dates(%\closure{$\ell_{4}$}{\emptyenv}%)),                         %\label{line:popend}%

SELECT o.o_orderkey,                                                            %\label{line:outerSFW}%
       o.o_orderstatus,
       dispatch_1(c.c_completion, o) AS completion
FROM   ORDERS o, COMPLETION c
WHERE  o.o_orderstatus = c.c_orderstatus;
  \end{lstlisting}
  \caption{\PLSQL{}~code of~\cref{fig:sql-h-o-PLSQL} after defunctionalization.}
  \label{fig:sql-f-o-PLSQL}
\end{figure}

\medskip\noindent
Applied to the example of~\cref{fig:sql-h-o-PLSQL} (order completion dates),
defunctionalization generates the output of~\cref{fig:sql-f-o-PLSQL}. The
resulting code executes on vanilla \PLSQL{} hosts; we show a Postgre\SQL{}\,9
dialect here, minor adaptations yield syntactic compatibility with Oracle.

\medskip\noindent
\PLSQL{} operates over typed tables and values and thus requires the
generation of \emph{typed closures}.  In the present example, we use
$\clos{\fun{t_{1}}{t_{2}}}$ to denote the type of closures that
represent functions of type $\fun{t_{1}}{t_{2}}$. (For now, $\clos{}$
is just a placeholder---\cref{sec:closure-representation} discusses
suitable relational implementations of this type.)   As expected, we find higher-order
function~\tpchcol{item}{dates} to accept and return values of such
types~$\clos{}$ (\cref{line:itemdates}).

Likewise, \PLSQL{} defunctionalization emits \emph{typed
dispatchers} \tpchcol{dispatch}{$i$} each of which implement
dynamic function invocation for closures of a particular
type:\footnote{Since~\PLSQL{} lacks parametric polymorphism, we
may assume that the~$t_{i}$ denote concrete types. Type specialization~\cite{MLtoAda} could
pave the way for a polymorphic variant of \PLSQL{}, one possible thread of future work.} the dispatcher associated with functions of
type~$\fun{t_{1}}{t_{2}}$ has the \PLSQL{} signature
\sql{FUNCTION($\clos{\fun{t_{1}}{t_{2}}}$,$t_{1}$)\;RETURNS\;$t_{2}$}. With this
typed representation come opportunities to improve efficiency.  We
turn to these in the next section.

\begin{figure}
  \centering\small
  \begin{littbl}
    \begin{tabular}{@{}|c|c|@{}}
      \tabname{2}{COMPLETION} \\
      \colhd{\tpchcol{c}{orderstatus}} & \colhd{\tpchcol{c}{completion}} \\
      \strut
      'F' & \closure{$\ell_{2}$}{\emptyenv} \\
      \strut
      'P' & \closure{$\ell_{3}$}{\emptyenv} \\
      \vrule width0pt height7pt depth7pt
      'O' & \closure{$\ell_{1}$}{%
              \vrule width0pt height5pt depth5pt
              \begin{environ}{c}
                \raisebox{-3pt}{\closure{$\ell_{4}$}{\emptyenv}}
              \end{environ}} \\
      \hline
    \end{tabular}
  \end{littbl}
  \caption{Table of functions: \sql{COMPLETION} holds closures of
    type~$\clos{\fun{\sql{ORDERS}}{\sql{DATE}}}$ in
    column~\tpchcol{c}{completion}.}
  \label{fig:sql-persistent-closures}
\end{figure}

\medskip\noindent
\textbf{Tables of Functions.}
After defunctionalization, functional values equate first-order closure
values.  This becomes apparent with a look at table~\sql{COMPLETION} after it
has been populated with three functions (in~\crefrange{line:popstart}{line:popend}
of~\cref{fig:sql-f-o-PLSQL}). Column~\tpchcol{c}{completion} holds the
associated closures (\cref{fig:sql-persistent-closures}). The closures with
labels~$\ell_{2}$ and~$\ell_{3}$ represent the function literals
in~\cref{line:litfun1,line:litfun2} of~\cref{fig:sql-h-o-PLSQL}: both are closed
and have an empty environment. Closure~$\ell_{1}$, representing the function
literal defined at~\cref{line:litfun3} of~\cref{fig:sql-h-o-PLSQL}, carries the
value of free variable~\sql{comp} which itself is a (date comparator)
function.  We thus end up with a nested closure.

Tables of functions may persist in the database like regular first-order
tables. To guarantee that closure labels and environment contents are
interpreted consistently when such tables are queried, update and query
statements need to be defunctionalized together, typically as part of the same
\PLSQL{} package~\cite[\S{}10]{OraclePLSQL} (\emph{whole-query
transformation}, see~\cref{sec:defunctionalization-for-xquery}).  Still,
query defunctionalization is restricted to operate in a \emph{closed world}:
the addition of new literal functions or named function references requires
the package to be defunctionalized anew.

\section{Representing (Nested) Closures}
\label{sec:closure-representation}

While the defunctionalization transformation nicely carries over to query
languages, we face the challenge to find closure representations that fit
query runtime environments.  Since we operate non-invasively, we need to
devise representations that can be expressed within the query language's data
model itself. (We might benefit from database engine adaptations
but such invasive designs are not in the scope of the present paper.)

Defunctionalization is indifferent to the exact method of closure construction
and elimination provided that the implementation can
\begin{compactenum}[(a)]
  \item discriminate on the code labels~$\ell$ and
  \item \label{item:any-value-in-env}
    hold any value of the language's data model in the environment.
    If the implementation is typed, we need~to
  \item ensure that all constructed closures for a given function
    type~$\fun{t_{1}}{t_{2}}$ share a common representation
    type~$\clos{\fun{t_{1}}{t_{2}}}$
    (cf.\ our discussion in \cref{sec:defunctionalization-plsql}).
\end{compactenum}
Since functions can assume the role of values,
(\labelcref{item:any-value-in-env}) implies that closures may be
\emph{nested}. We encountered nested closures of depth~2
in~\cref{fig:sql-persistent-closures} where the environment of
closure~$\ell_{1}$ holds a closure labeled~$\ell_{4}$.  For
particular programs, the nesting depth may be unbounded, however.  The associative
map example of~\cref{sec:first-class-functions} creates closures of the form
\medskip
\newcommand{\talldash}[3]{%
  \setbox0\vbox to #1{
    {\color{closure}\leaders\vbox{\vskip1pt\hrule height 1pt width.5pt}\vfill}}%
  \dp0=#2\ht0=#3
  \box0
}
\begin{equation}
  \closure{$\ell_{1}$}{
    \begin{environ}{c}
      $k_{1}$\,\talldash{21pt}{8pt}{13pt}\,%
      $v_{1}$\,\talldash{21pt}{8pt}{13pt}\,%
        \closure{$\ell_{1}$}{
          \begin{environ}{c}
            $k_{2}$\,\talldash{18pt}{6.5pt}{11.5pt}\,%
            $v_{2}$\,\talldash{18pt}{6.5pt}{11.5pt}\,%
            \closure{$\ell_{1}$}{
                \begin{environ}{c}
                  $\cdots$ \vrule height10pt depth5pt width0pt
                  \closure{$\ell_{1}$}{
                    \begin{environ}{c}
                      $k_{n}$\,\talldash{12pt}{4pt}{8pt}\,%
                      $v_{n}$\,\talldash{12pt}{4pt}{8pt}\,%
                      \closure{$\ell_{3}$}{\emptyenv}
                    \end{environ}
                  }
                \end{environ}
              }
          \end{environ}
        }
    \end{environ}
  }
  \tag{$\ast$}
  \label{eq:nested-closure}
\end{equation}

\medskip\noindent
where the depth is determined by the number $n$ of key/value pairs
$(k_{i},v_{i})$ stored in the map.

Here, we discuss closure implementation variants in terms of representation
functions~$\repr\llbracket\cdot\rrbracket$ that map closures to regular
language constructs. We also point out several refinements.

\subsection{\XQuery{}: Tree-Shaped Closures}
\label{sec:closure-representation-xquery}

For \XQuery{}, one representation that equates closure construction with
\XML{} element construction is given in~\cref{fig:closure-construction-xquery}. A
closure with label~$\ell$ maps to an outer element with tag~$\ell$ that
holds the environment contents in a sequence of~\xquery{env} elements.
In the environment, atomic items are tagged with their dynamic type such
that closure elimination can restore value and type (note the
calls to function~\xquery{wrap()} and its definition in~\cref{fig:wrap-unwrap-items}):
item~\xquery{1} of type \xquery{xs:integer} is held
as~\xquery{<atom><integer>1</integer></atom>}. Item sequences
map into sequences of their wrapped items, \XML{} nodes are not wrapped at
all.

\begin{figure}[t]
  \centering\small
  $
  \begin{array}{@{}rcl@{}}
    \repr\llbracket\closure{$\ell$}{%
      \begin{environ}{eec}
        $x_{1}$ & $\cdots$ & $x_{n}$
      \end{environ}}\rrbracket
    & = &
    \begin{array}[t]{@{}l@{\,}l@{}}
    |element|\,\ell\,|{| & |element|\,|env|\,|{|\,\repr\llbracket x_{1}\rrbracket\,|}||,|\dots|,| \\
                         & |element|\,|env|\,|{|\,\repr\llbracket x_{n}\rrbracket\,|}|\,|}|
    \end{array}
    \\
    \repr\llbracket\closure{$\ell$}{\emptyenv}\rrbracket
    & = &
    |element|\,\ell\,|{}|
    \\
    \repr\llbracket x\rrbracket
    & = &
    \xquery{wrap($x$)}
  \end{array}
  $
  \caption{\XQuery{} closure representation in terms of \XML{} fragments.
    Function~\xquery{wrap()} is defined in~\cref{fig:wrap-items}.}
  \label{fig:closure-construction-xquery}
\end{figure}

\begin{figure}[t]%
  \centering\small%
  \begin{SubFloat}{\label{fig:wrap-items}}%
  \begin{lstlisting}[boxpos=t]
declare function wrap($xs)
{
 for $x in $xs return
  typeswitch ($x)
   case xs:anyAtomicType
    return wrap-atom($x)
   case attribute(*)
    return element attr {%\,%$x%\,%}
   default return $x
};

%%
  \end{lstlisting}%
  \end{SubFloat}%
  \hfill\vrule\hfill
  \begin{SubFloat}{\label{fig:wrap-atomic-item}}%
  \begin{lstlisting}[boxpos=t,numbers=none]
declare function wrap-atom($a)
{
 element atom {
  typeswitch ($a)
   case xs:integer
    return element integer {%\,%$a%\,%}
   case xs:string
    return element string {%\,%$a%\,%}
     %\textnormal{[\dots more atomic types\dots]}%
   default return element any {%\,%$a%\,%}
  }
};
  \end{lstlisting}%
  \end{SubFloat}%
  \caption{Preserving value and dynamic type of environment contents through
    wrapping.}
  \label{fig:wrap-unwrap-items}
\end{figure}

Closure elimination turns into an \XQuery{}~\xquery{typeswitch()} on the
outer tag name while values in the environment are accessed via
\XPath{} \xquery{child} axis steps (\cref{fig:closure-elimination-xquery}). Auxiliary function~\xquery{unwrap()}
(obvious, thus not shown) uses the type tags to restore the original
atomic items held in the environment.

\medskip\noindent
In this representation, closures nest naturally.  If we apply
$\repr\llbracket\cdot\rrbracket$ to the closure~\labelcref{eq:nested-closure}
that resulted from key/value map construction,
we obtain the \XML{} fragment of~\cref{fig:map-xml-fragment} whose nested shape directly
reflects that of the input closure.

\medskip\noindent
\textbf{Refinements.} The above closure representation builds on
inherent strengths of the underlying \XQuery{} processor---element
construction and tree navigation---but has its shortcomings: \XML{}
nodes held in the environment lose their original tree context due to
\XQuery{}'s copy semantics of node construction. If this
affects the defunctionalized queries, an environment representation
based on \emph{by-fragment} semantics~\cite{XRPC}, preserving
document order and ancestor context, is a viable alternative.

Further options base on \XQuery{}'s other aggregate data type: the item
sequence: closures then turn into non-empty sequences of type~\xquery{item()+}.
While the head holds label~$\ell$, the tail can hold the environment's contents:
\xquery{($\ell$,$x_{1}$,$\dots$,$x_{n}$)}. In this representation,
neither atomic items nor nodes require wrapping as value, type,
and tree context are faithfully preserved. Closure elimination
accesses the~$x_{i}$ through simple positional lookup into the
tail. Indeed, we have found this implementation option to perform
particularly well (\cref{sec:experiments}). Due to \XQuery{}'s
implicit sequence flattening, this variant requires additional
runtime effort in the presence of sequence-typed~$x_{i}$ or
closure nesting, though (techniques for the flat representation of
nested sequences apply~\cite{Dremel}).

Lastly, invasive approaches may build on engine-internal support for
aggregate data structures. Saxon~\cite{Dremel}, for example, implements
an appropriate tuple structure that can serve to represent
closures.\footnote{\url{http://dev.saxonica.com/blog/mike/2011/07/#000186}}

\begin{figure}
  \small\centering
  \lstset{language=xquery,numbers=none}
  \begin{tabular}[t]{@{}r@{\,}c@{}l@{}}
     $
     \begin{array}{@{}l@{}}
       \meta{case}~e_{1}~\meta{of} \\
       \quad \vdots \\[1ex]
       \quad \closure{$\ell_{}$}{\begin{environ}{eec}
                       $\vardollar{}v_{1}$ & $\cdots$ & $\vardollar{}v_{n}$
                     \end{environ}} \Rightarrow e_{2}
     \end{array}
     $
     &
     $\rightsquigarrow$
     &
    \begin{minipage}{0.6\linewidth}
    \begin{lstlisting}
typeswitch (%$e_{1}$%)
    %\raisebox{-2pt}{\smash{$\vdots$}}%
  case element(%$\ell$%) return
    let $env := %$e_{1}$%/env
    let $%$v_{1}$% := unwrap($env[1]/node())
                 %\raisebox{-2pt}{\smash{$\vdots$}}%
    let $%$v_{\phantom{1}\mathllap n}$% := unwrap($env[%$n$%]/node())
    return %$e_{2}$%
    \end{lstlisting}
    \end{minipage}
    \\[-2ex]
  \end{tabular}%
  \caption{\XQuery{} closure elimination: \xquery{typeswitch()} discriminates
      on the label, axis steps access the environment.}
  \label{fig:closure-elimination-xquery}
\end{figure}

\begin{figure}
  \centering\small
    \begin{minipage}{0.75\linewidth}
    \begin{lstlisting}[language=XML,basicstyle=\color{black!60}\ttfamily,lineskip=-10pt]
<%\color{black}$\ell_{1}$%>
  <env><atom><%\smash{$t_{\var{key}}$}%>%\color{black}$k_{1}$%</%\smash{$t_{\var{key}}$}%></atom></env>
  <env><atom><%\smash{$t_{\var{val}}$}%>%\color{black}$v_{1}$%</%\smash{$t_{\var{val}}$}%></atom></env>
  <env>
    <%\color{black}$\ell_{1}$%>
      <env><atom><%\smash{$t_{\var{key}}$}%>%\color{black}$k_{2}$%</%\smash{$t_{\var{key}}$}%></atom></env>
      <env><atom><%\smash{$t_{\var{val}}$}%>%\color{black}$v_{2}$%</%\smash{$t_{\var{val}}$}%></atom></env>
      <env>
        %\vrule height6pt depth0pt width0pt\smash{$\cdots$}%
        <%\color{black}$\ell_{1}$%>
          <env><atom><%\smash{$t_{\var{key}}$}%>%\color{black}$k_{n}$%</%\smash{$t_{\var{key}}$}%></atom></env>
          <env><atom><%\smash{$t_{\var{val}}$}%>%\color{black}$v_{n}$%</%\smash{$t_{\var{val}}$}%></atom></env>
          <env><%\color{black}$\ell_{3}$%/></env>
        </%\color{black}$\ell_{1}$%>
        %\vrule height5pt depth1pt width0pt\smash{$\cdots$}%
      </env>
    </%\color{black}$\ell_{1}$%>
  </env>
</%\color{black}$\ell_{1}$%>
    \end{lstlisting}
    \end{minipage}
  \caption{\XML{} representation of the nested closure~\labelcref{eq:nested-closure}.
    $t_{\var{key}}$ and $t_{\var{val}}$ denote the types of keys and values, respectively.}
  \label{fig:map-xml-fragment}
\end{figure}

\subsection{\PLSQL{}: Typed Closures}
\label{sec:closure-representation-plsql}

Recall that we require a fully typed closure representation to meet the
\PLSQL{} semantics (\cref{sec:defunctionalization-plsql}). A direct representation
of closures of, in general, unbounded depths would call for a \emph{recursive}
representation type. Since the \PLSQL{} type system reflects the flat relational
data model, recursive types are not permitted, however.

Instead, we represent closures as \emph{row values}, built by
constructor~\sql{ROW()}, \emph{i.e.}, native aggregate record structures
provided by \PLSQL{}. Row values are first-class citizens in \PLSQL{} and, in
particular, may be assigned to variables, can contain nested row values, and
may be stored in table cells (these properties are covered by
feature~\emph{S024} ``support for enhanced structured types'' of the
\SQL{}:1999 standard~\cite{SQL:1999}).

\medskip\noindent
\cref{fig:closure-construction-sql} defines function~$\repr\llbracket\cdot\rrbracket$
that implements a row value-based representation.
A closure
\closure{$\ell$}{
  \begin{environ}{eec}
    $x_1$ & $\cdots$ & $x_n$
  \end{environ}
}
of type~$\clos{\fun{t_{1}}{t_{2}}}$ maps to the
expression~$\sql{ROW($\ell$,$\gamma$)}$.  If the environment is non-empty,
$\repr\llbracket\cdot\rrbracket$ constructs an additional row to hold the
environment contents.  This row, along with key~$\gamma$ is then appended to
binary table~$\sql{ENV}_{\fun{t_{1}}{t_{2}}}$ which collects the environments
of all functions of type~$\fun{t_{1}}{t_{2}}$.  Notably, we represent
non-closure values~$x$ as is ($\repr\llbracket x\rrbracket = x$), saving the
program to perform $\xquery{wrap()}$/$\xquery{unwrap()}$ calls at runtime.

\begin{figure}[t]
  \centering\small
  $
  \begin{array}{@{}rcl@{}}
    \repr\llbracket\closure{$\ell$}{%
      \begin{environ}{eec}
        $x_{1}$ & $\cdots$ & $x_{n}$
      \end{environ}}\rrbracket
    & = &
    \sql{ROW($\ell$,$\gamma$)}
    \hskip6mm\vrule
    \text{%
    \begin{littbl}%
      \begin{tabular}{@{}|c|c|@{}}
        \tabname{2}{$\boldmath\sql{ENV}_{\fun{t_{1}}{t_{2}}}$} \\
        \colhd{\sql{id}} & \colhd{\sql{env}} \\
        $\vdots$ & $\vdots$ \\
        $\gamma$ & \sql{ROW($\repr\llbracket x_{1}\rrbracket$,$\dots$,$\repr\llbracket x_{n}\rrbracket$)}\\
        \hline
      \end{tabular}
    \end{littbl}}
    \\
    \repr\llbracket\closure{$\ell$}{\emptyenv}\rrbracket
    & = &
    \sql{ROW($\ell$,NULL)}
    \\
    \repr\llbracket x\rrbracket & = & x
  \end{array}
  $
  \caption{Relational representation for closures, general approach ($\gamma$
    denotes an arbitrary but unique key value).}
  \label{fig:closure-construction-sql}
\end{figure}

\begin{figure}[t]
  \centering\small
  \begin{littbl}
    \begin{tabular}{@{}|c|c|@{}}
      \tabname{2}{$\boldmath\sql{ENV}_{\fun{t_{\var{key}}}{t_{\var{val}}}}$} \\
      \colhd{\sql{id}} & \colhd{\sql{env}} \\
      \strut
      $\gamma_{n}$   & \sql{ROW($k_{1}$,$v_{1}$,ROW($\ell_{1}$,$\gamma_{n-1}$))} \\
      $\gamma_{n-1}$ & \sql{ROW($k_{2}$,$v_{2}$,ROW($\ell_{1}$,$\gamma_{n-2}$))} \\[-3pt]
      $\vdots$       & $\vdots$ \\
      $\gamma_{1}$   & \sql{ROW($k_{n}$,$v_{n}$,ROW($\ell_{3}$,NULL))}
      \\
      \hline
    \end{tabular}
  \end{littbl}
  \caption{Environment table built to represent closure~\labelcref{eq:nested-closure}.}
  \label{fig:environment-table}
\end{figure}

This representation variant yields a flat relational encoding
regardless of closure nesting depth.  \cref{fig:environment-table} depicts
the table of environments that results from encoding
closure~\labelcref{eq:nested-closure}.
The overall top-level closure is represented by~\sql{ROW($\ell_{1}$,$\gamma_{n}$)}:
construction proceeds inside-out with a new outer closure layer added whenever
a key/value pair is added to the map. This representation of closure
environments matches well-known relational encodings of tree-shaped data
structures~\cite{EdgeApproach}.

\medskip\noindent
\textbf{Environment Sharing.}
\sql{ENV} tables create opportunities for environment sharing. This
becomes relevant if function literals are evaluated under invariable bindings
(recall our discussion of function~\xquery{group-by}
in~\cref{fig:xquery-h-o-group-by}).  A simple, yet dynamic implementation of
environment sharing is obtained if we alter the behavior of
$\repr\llbracket\closure{$\ell$}{%
  \begin{environ}{eec}
    $x_{1}$ & $\cdots$ & $x_{n}$
  \end{environ}}\rrbracket$:
when the associated~\sql{ENV} table already carries an environment of the same
contents under a key~$\gamma$, we return~\sql{ROW($\ell$,$\gamma$)} and do not
update the table---otherwise a new environment entry is appended as described before.
Such \emph{upsert} operations are a native feature of recent \SQL{} dialects
(cf.\ \sql{MERGE}~\cite[\S{}14.9]{SQL:1999}) and benefit if column~\sql{env} of the~\sql{ENV}
table is indexed. The resulting many-to-one relationship between closures and
environments closely resembles the space-efficient \emph{safely linked closures}
as described by Shao and Appel in~\cite{LinkedClosures}. We return to environment sharing
in~\cref{sec:experiments}.

\medskip\noindent
\textbf{Closure Inlining.}
Storing environments separately from their closures also incurs an overhead
during closure elimination, however.  Given a closure
encoding~\sql{ROW($\ell$,$\gamma$)} with $\gamma \neq \sql{NULL}$,
the dispatcher
\begin{inparaenum}[(1)]
\item discriminates on~$\ell$, \emph{e.g.}, via
  \PLSQL{}'s~\sql{CASE$\cdots$WHEN$\cdots$END\,CASE},
  then
\item accesses the environment through an~\sql{ENV} table lookup with
  key~$\gamma$.
\end{inparaenum}

With typed closures, the representation types~$\clos{\fun{t_{1}}{t_{2}}}$
are comprised of (or: depend on) typed environment contents. For the large
class of programs---or parts thereof---which nest closures to a statically known,
limited depth, these representation types will be \emph{non-recursive}.
Below, the type dependencies for the examples of~\cref{fig:xquery-h-o-map,fig:sql-h-o-PLSQL}
are shown on the left and right, respectively
(read~\tikz[baseline=-2pt]\draw[->,gray,bend left=25] (0,0) to (0.5,0); as
``has environment contents of type''):
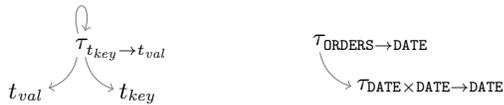
$$
\begin{array}{c@{\hskip2cm}c}
  \begin{tikzpicture}[->,outer sep=1pt,inner sep=1pt,node distance=4mm]
    \tikzstyle{depend}=[gray]
    \node (t) {$\clos{\args{\fun{t_{\var{key}}}{t_{\var{val}}}}}$};
    \node (key) [below right=of t]  {$t_{\var{key}}$};
    \node (val) [below left=of t] {$t_{\var{val}}$};
    \path[depend] (t) edge [loop above] (t);
    \path[depend] (t) edge [bend right] (key);
    \path[depend] (t) edge [bend left] (val);
  \end{tikzpicture}
  &
  \begin{tikzpicture}[baseline=-25pt,->,outer sep=1pt,inner sep=1pt,node distance=4mm]
    \tikzstyle{depend}=[gray]
    \node (t1) {$\clos{\args{\fun{\sql{ORDERS}}{\sql{DATE}}}}$};
    \node (t2) [below right=of t1] {$\clos{\fun{\sql{DATE} \times \sql{DATE}}{\sql{DATE}}}$};
    \path[depend] (t1) edge [bend right] (t2.west);
  \end{tikzpicture}
\end{array}
$$
Note how the loop on the left coincides with the recursive shape of
closure~\labelcref{eq:nested-closure}. If these dependencies are acyclic (as
they are for the order completion date example), environment contents may be
kept directly with their containing closure: separate~\sql{ENV} tables are not
needed and lookups are eliminated entirely.
\cref{fig:inline-closure-construction-sql} defines a variant of~$\repr\llbracket\cdot\rrbracket$
that implements this \emph{inlined} closure representation.  With this
variant, we obtain
$\repr\llbracket\mskip2mu
  \closure{$\ell_{1}$}{%
    \vrule width0pt height5pt depth3pt
    \begin{environ}{c}
      \closure{$\ell_{4}$}{\emptyenv}
    \end{environ}}\mskip2mu\rrbracket = \sql{ROW($\ell_{1}$,ROW($\ell_{4}$,NULL))}$
(see \cref{fig:sql-persistent-closures}).

\begin{figure}[t]
  \centering\small
  $
  \begin{array}{@{}rcl@{}}
    \repr\llbracket\closure{$\ell$}{%
      \begin{environ}{eec}
        $x_{1}$ & $\cdots$ & $x_{n}$
      \end{environ}}\rrbracket
    & = &
    \sql{ROW($\ell$,ROW($\repr\llbracket x_{1}\rrbracket$,$\dots$,$\repr\llbracket x_{n}\rrbracket$))}
    \\[2pt]
    \repr\llbracket\closure{$\ell$}{\emptyenv}\rrbracket
    & = &
    \sql{ROW($\ell$,NULL)}
    \\[2pt]
    \repr\llbracket x\rrbracket & = & x
  \end{array}
  $
  \caption{Relational representation of closures with fixed nesting depth:
    environment contents inlined into closure.}
  \label{fig:inline-closure-construction-sql}
\end{figure}

We quantify the savings that come with closure inlining in the upcoming
section.


\section{Does it Function? (Experiments)}
\label{sec:experiments}

Adding native support for first-class functions to a first-order
query processor calls for disruptive changes to its data model and
the associated set of supported operations. With defunctionalization
and its non-invasive source transformation, these changes
are limited to the processor's front-end (parser, type checker, query
simplification). Here, we explore this positive aspect but also quantify the
performance penalty that the non-native defunctionalization approach incurs.

\medskip\noindent
\textbf{\XQuery~3.0 Test Suite.}
Given the upcoming \XQuery{}~3.0 standard, defunctionalization can help to carry
forward the significant development effort that has been put into \XQuery~1.0
processors. To make this point, we subjected three such
processors---\Oracle{}~(release~11.1)~\cite{OracleXQuery}, Berkeley~DB
\XML{}~2.5.16~\cite{BerkeleyDBXML} and Sedna~3.5.161~\cite{Sedna}---to relevant
excerpts of the W3C~\XQuery{}~3.0 Test Suite (XQTS).\footnote{A pre-release is
  available
  at~\url{http://dev.w3.org/cvsweb/2011/QT3-test-suite/misc/HigherOrderFunctions.xml}.}
All three engines are \emph{database-supported} \XQuery{}
processors; native support for first-class functions would require substantial
changes to their database kernels.

Instead, we fed the XQTS queries into a stand-alone preprocessor
that implements the defunctionalization transformation as described
in~\cref{sec:defunctionalization}.
The test suite featured, \emph{e.g.},
\begin{compactitem}
  \item named references to user-defined and built-in functions, literal
    functions, sequences of functions, and
  \item higher-order functions accepting and returning functions.
\end{compactitem}
All three systems were able to successfully pass these tests.

\medskip\noindent
\textbf{Closure Size.}
We promote a function-centric query style in this work, but ultimately all
queries have to be executed by data-centric database query engines.
Defunctionalization implements this transition from functions to data,
\emph{i.e.}, closures, under the hood. This warrants a look at closure size.

Turning to the \XQuery{} grouping example of~\cref{fig:xquery-h-o-group-by}
again, we see that the individual groups in the sequence returned by
\xquery{group-by} are computed \emph{on-demand}: a group's members will be
determined only once its function is applied (\xquery{\$g()}
in~\cref{line:arg-less-call}). Delaying the evaluation of expressions by
wrapping them into (argument-less) functions is another useful idiom available
in languages with first-class functions~\cite{Thunking}, but there are
implications for closure size: each group's closure captures the environment
required to determine its group members. Besides~\xquery{\$key}
and~\xquery{\$k}, each environment includes the contents of free
variable~\xquery{\$seq} (the input sequence) such that the overall closure
space requirements are in $O(g \cdot \lvert\xquery{\$seq}\rvert)$ where~$g$
denotes the number of distinct groups.  A closure representation that allows
the sharing of environments (\cref{sec:closure-representation-plsql}) would bring
the space requirements down to $O(\lvert\xquery{\$seq}\rvert)$ which marks the
minimum size needed to partition the sequence~\xquery{\$seq}.

\begin{figure}[t]
  \centering\small
  \begin{lstlisting}[language=xquery]
declare function group-by($seq as item()*,
                          $key as function(item()*) as item()*)
  as (function() as item()*)*
{
  let $keys := for $x in $seq return $key($x)
  for $k in distinct-values($keys)
  let $group := $seq[$key(.) = $k]
    return                           %\smash[tb]{$\biggr]$}~\textnormal{changed from~\cref{fig:xquery-h-o-group-by}}%
      function() { $group }              %\label{line:free-var-group}%
};

let $fib := (0,1,1,2,3,5,8,13,21,34)
for $g in group-by($fib, function($x) { $x mod 2 })
return
  element group { $g() }
  \end{lstlisting}
  \caption{Hoisting invariant computation out of the body of the literal
    function at~\cref{line:free-var-group} affects closure size.}
  \label{fig:xquery-h-o-opt-group-by}
\end{figure}

Alternatively, in the absence of sharing, evaluating the
expression~\xquery{\$seq[\$key(.)\,=\,\$k]}
outside the wrapping function computes groups \emph{eagerly}.
\cref{fig:xquery-h-o-opt-group-by} shows this alternative
approach in which the bracketed part has been changed
from~\cref{fig:xquery-h-o-group-by}. A group's closure now only
includes the group's members (free variable~\xquery{\$group},
\cref{line:free-var-group} in~\cref{fig:xquery-h-o-opt-group-by}) and the
overall closure sizes add up to $O(\lvert\xquery{\$seq}\rvert)$ as desired.
Closure size thus should be looked at with care during query formulation---such
``space leaks'' are not specific to the present approach,
however~\cite{SpaceLeaks}.

\begin{figure}
  \centering\small
    \begin{minipage}{0.75\linewidth}
    \begin{lstlisting}[language=XML,basicstyle=\color{black!60}\ttfamily,lineskip=-10pt]
<map>
  <entry>
    <key><atom><%\smash{$t_{\var{key}}$}%>%\color{black}$k_{1}$%</%\smash{$t_{\var{key}}$}%></atom></key>
    <val><atom><%\smash{$t_{\var{val}}$}%>%\color{black}$v_{1}$%</%\smash{$t_{\var{val}}$}%></atom></val>
  </entry>
  <entry>
    <key><atom><%\smash{$t_{\var{key}}$}%>%\color{black}$k_{2}$%</%\smash{$t_{\var{key}}$}%></atom></key>
    <val><atom><%\smash{$t_{\var{val}}$}%>%\color{black}$v_{2}$%</%\smash{$t_{\var{val}}$}%></atom></val>
  </entry>
  %\vrule height5pt depth1pt width0pt\smash{$\cdots$}%
  <entry>
    <key><atom><%\smash{$t_{\var{key}}$}%>%\color{black}$k_{n}$%</%\smash{$t_{\var{key}}$}%></atom></key>
    <val><atom><%\smash{$t_{\var{val}}$}%>%\color{black}$v_{n}$%</%\smash{$t_{\var{val}}$}%></atom></val>
  </entry>
</map>
    \end{lstlisting}
    \end{minipage}
  \caption{Key-value map representation generated by the first-order
    code of~\cref{fig:xquery-f-o-element-map} (compare with the closure of~\cref{fig:map-xml-fragment}).}
  \label{fig:map-f-o-xml-fragment}
\end{figure}

\medskip\noindent
With defunctionalization, queries lose functions but gain data.  This does not
imply that defunctionalized queries use inappropriate amounts of space, though.
In our experiments we have found function-centric queries to implicitly generate
closures whose size matches those of the data structures that are explicitly
built by equivalent first-order formulations.

To illustrate, recall the two \XQuery{} map variants of~\cref{sec:first-class-functions}.
Given~$n$ key/value pairs $(k_i,v_i)$, the function-centric variant
of~\cref{fig:xquery-h-o-map} implicitly constructs the nested
closure shown in~\cref{fig:map-xml-fragment}: a non-empty map of $n$~entries
will yield a closure size of $10\cdot n$~\XML{} nodes. In comparison, the
first-order map variant of~\cref{fig:xquery-f-o-element-map} explicitly builds a key/value
list of similar size, namely~$1 + 9\cdot n$ nodes (\cref{fig:map-f-o-xml-fragment}).
Further, key lookups in the map incur almost identical \XPath{} navigation
efforts in both variants, either through closure elimination or, in the
first-order case, the required calls to~\xquery{map:get}.

\medskip\noindent
\textbf{Native vs.\ Dispatched Function Calls.}
As expected, the invocation of functions through closure label discrimination
by dispatchers introduces measurable overhead if compared to
native function calls.\footnote{Remember that this overhead only applies to
dynamic function calls---static calls are still performed natively.} To quantify
these costs, we performed experiments in which $10^{6}$ native and dispatched
calls were timed.  We report the averaged wall-clock times of 10~runs measured
on a Linux host, kernel version~3.5, with Intel Core~i5 CPU (2.6\,GHz) and
8\,GB of primary memory.

\begin{table}[t]
  \centering\small
  \subfloat[Unary \PLSQL{} function.
    \label{tab:sql-dyn-call}]{%
    \begin{tabular}{@{}rcc@{}}
      \toprule
      & \textbf{Oracle} & \textbf{Postgre\SQL} \\
      \cmidrule(lr){2-2}\cmidrule(l){3-3}
      native            & 10\,500 & 2\,414 \\   
      \xquery{dispatch} & 11\,860 & 8\,271                           
      \\
      \bottomrule
    \end{tabular}}
  \hfill
  \subfloat[Literal \XQuery{} function.
    \label{tab:xquery-dyn-call}]{%
    \begin{tabular}{@{}rcc@{}}
      \toprule
                        & \textbf{BaseX} & \textbf{Saxon} \\
      \cmidrule(lr){2-2}\cmidrule(l){3-3}
      native            & 394            & 1\,224 \\
      \xquery{dispatch} & 448            & \phantom{1\,}755
      \\
      \bottomrule
    \end{tabular}}
  \vskip1pt
  \caption{Performing 10$^\mathbf{6}$ invocations of closed functions
    (native vs.\ \sql{dispatch}ed calls). Wall-clock time measured in~ms.}
  \label{tab:sql-xquery-dyn-call}
\end{table}

Both, function invocation itself and closure manipulation contribute
to the overhead. To assess their impact separately, a first round
of experiments invoked \emph{closed} functions (empty environment).
\cref{tab:sql-dyn-call} documents the cost of a dispatched \PLSQL{}
function call---\emph{i.e.}, construction of an empty closure, static
call to the \sql{dispatch} function, closure label discrimination,
static call to a surrogate function. While dispatched function
calls minimally affect \Oracle{} performance---hinting at a remarkably efficient
implementation of its \PLSQL{} interpreter---the cost is
apparent in Postgre\SQL{}~9.2 (factor~$3.5$). In the \XQuery{}
case, we executed the experiment using~BaseX~7.3~\cite{BaseX} and
Saxon~9.4~\cite{Saxon}---both engines provide built-in support for
\XQuery~3.0 and thus allow a comparison of the costs of a native
versus a defunctionalized implementation of first-class functions.
BaseX, for example, employs a Java-based implementation of closure-like structures
that refer to an expression tree and a variable environment.
For the dynamic invocation of a closed literal function, BaseX shows a moderate
increase of $14\,\%$ (\cref{tab:xquery-dyn-call}) when dispatching is used.
For Saxon, we see a
\emph{decrease} of~$38\,\%$ from which we conclude that Saxon implements
static function calls (to~\xquery{dispatch} and the surrogate function
in this case) considerably more efficient than dynamic calls.
The resulting performance advantage of defunctionalization has also been reported by
Tolmach and Oliva~\cite{MLtoAda}.

\begin{table}
  \centering\small
  \begin{tabular}{@{}rrrrrrr@{}}
    \toprule
    & \multicolumn{3}{c}{\textbf{BaseX}} & \multicolumn{3}{c}{\textbf{Saxon}} \\
    \#\kern1pt free variables
      & \multicolumn{1}{c}{\scriptsize\textbf{1}} & \multicolumn{1}{c}{\scriptsize\textbf{5}} & \multicolumn{1}{c}{\scriptsize\textbf{10}}
      & \multicolumn{1}{c}{\scriptsize\textbf{1}} & \multicolumn{1}{c}{\scriptsize\textbf{5}} & \multicolumn{1}{c}{\scriptsize\textbf{10}} \\
    \cmidrule(lr){2-4}\cmidrule(l){5-7}
    native   & 402    & 396    & 467     & 1\,144 & 1\,451 & 1\,725  \\[1ex]
    node     & 2\,132 & 7\,685 & 14\,535 & 2\,133 & 7\,347 & 12\,992 \\
    sequence & 743    & 1\,527 & 2\,485  & 854    & 1\,526 & 2\,350
    \\
    \bottomrule
  \end{tabular}
  \vskip5pt
  \caption{10$^\mathbf{6}$ invocations and elimination of closures of varying size
    (1/5/10~free variables). Wall-clock time measured in~ms.}
  \label{tab:xquery-dyn-call-closures}
\end{table}

\begin{table}
  \begin{narrow}{-2mm}{-2mm}
  \centering\small
  \begin{tabular}{@{}r>{\scriptsize}r>{\ttfamily}lrr@{}}
    \toprule
    \textbf{\#\,Calls} &
    \small{\textbf{Line}} &
    \textnormal{\textbf{Query/Function}} &
    \multicolumn{1}{c}{\textbf{\sql{ENV}}} &
    \multicolumn{1}{c@{}}{\textbf{Inline}}
    \\
    \cmidrule(r){1-1}\cmidrule(lr){2-3}\cmidrule(lr){4-4}\cmidrule(l){5-5}
              1 & \labelcref{line:outerSFW}  & \sql{SELECT\,\tpchcol{o}{orderkey,}}$\cdots$                              & 47\,874 & 40\,093 \\
    1\,500\,000 & \labelcref{line:dispatch1} & \textSFii\textSFx\sql{\tpchcol{dispatch}{1}()}                            & 44\,249 & 35\,676 \\
       732\,044 & \labelcref{line:l1}        & ~~\textSFviii\textSFx\sql{$\ell_{1}$()}                                   & 22\,748 & 22\,120 \\
       732\,044 & \labelcref{line:innerSFW}  & ~~\textSFxi~\textSFii\textSFx\sql{SELECT\,\tpchcol{dispatch}{2}($\cdots$} &  9\,270 &  9\,363 \\
       732\,044 & \labelcref{line:dispatch2} & ~~\textSFxi~~~\textSFii\textSFx\sql{\tpchcol{dispatch}{2}()}              &  3\,554 &  3\,450 \\
       729\,413 & \labelcref{line:l2}        & ~~\textSFviii\textSFx\sql{$\ell_{2}$()}                                   &  2\,942 &  2\,856 \\
        38\,543 & \labelcref{line:l3}        & ~~\textSFii\textSFx\sql{$\ell_{3}$()}                                     &     155 &     149 \\
    \bottomrule
  \end{tabular}
  \end{narrow}
  \caption{Profiles for the \PLSQL{} program of~\cref{fig:sql-f-o-PLSQL}:
    environment tables vs.\ closure inlining. Averaged cumulative time
    measured in ms. Line numbers refer to~\cref{fig:sql-f-o-PLSQL}.}
  \label{tab:profile-plsql}
\end{table}

In a second round of experiments, we studied the dynamic invocation of \XQuery{}
functions that access~1, 5, or 10~free variables of
type~\xquery{xs:integer}. The defunctionalized implementation shows the expected
overhead that grows with the closure size
(see~\cref{tab:xquery-dyn-call-closures}): the dispatcher needs to extract and
unwrap~1, 5, or 10~environment entries from its closure argument~\xquery{\$clos}
before these values can be passed to the proper surrogate function
(\cref{sec:defunctionalization}). As anticipated
in~\cref{sec:closure-representation-xquery}, however, a sequence-based representation of
closures can offer a significant improvement over the \XML{} node-based
variant---both options are shown in~\cref{tab:xquery-dyn-call-closures} (rows
``node'' \emph{vs.}\ ``sequence''). If this option is applicable, the saved node
construction and \XPath{} navigation effort allows the defunctionalized
invocation of non-closed functions perform within a factor of~$1.36$ (Saxon)
or~$5$ (BaseX) of the native implementation.

\medskip\noindent
\textbf{Environment Tables vs.\ Closure Inlining.}
Zooming out from the level of individual function calls, we assessed the
runtime contribution of dynamic function calls and closure elimination in the
context of a complete \PLSQL{} program (\cref{fig:sql-f-o-PLSQL}). To this
end, we recorded time profiles while the program was evaluated against a
\TPCH{} instance of scale factor~$1.0$ (the profiles are based on
Postgre\SQL{}'s \textmtt{pg\_stat\_statements} and
\textmtt{pg\_stat\_user\_functions} views~\cite{PostgreSQL}).
\cref{tab:profile-plsql} shows the cumulative times (in ms) over all query and
function invocations: one evaluation of~\sql{\tpchcol{dispatch}{1}()},
including the queries and functions it invokes, takes $\nicefrac{\displaystyle
44\,429~\text{ms}}{\displaystyle 1\,500\,000} \approx 0.03~\text{ms}$ on
average (column~\sql{ENV}). The execution time of the top-level~\sql{SELECT}
statement defines
the overall execution time of the program. Note that the cumulative times do
not add up perfectly since the inevitable \PLSQL{} interpreter overhead and
the evaluation of built-in functions are not reflected in these profiles.

Clearly, \sql{\tpchcol{dispatch}{1}()} dominates the profile as it embodies
the core of the configurable completion date computation. For more
than~$50\,\%$ of the overall $1\,500\,000$~orders, the dispatcher needs to
eliminate a closure of type~$\clos{\fun{\sql{ORDERS}}{\sql{DATE}}}$ and
extract the binding for free variable~\sql{comp} from its environment before
it can invoke surrogate function~\sql{$\ell_{1}$()}. According
to~\cref{sec:closure-representation-plsql}, closure inlining is applicable
here and column~\textbf{Inline} indeed shows a significant reduction of execution
time by~$18\,\%$
(\sql{\tpchcol{dispatch}{2}()} does not benefit since it exclusively processes
closures with empty environments.)

\medskip\noindent
\textbf{Simplifications.}
A series of simplifications help to further reduce the cost
of queries with closures:
\begin{compactitem}
\item Identify $\closure{$\ell$}{\emptyenv}$ and $\ell$ (do not build closures with
  empty environment).  This benefits dynamic calls to closed and built-in functions.
\item If~$\Dispatch{n}$ is a singleton set, \xquery{dispatch\_$n$} becomes superfluous
  as it is statically known which $\meta{case}$ branch will be taken.
\item When constructing~$\closure{$\ell$}{%
  \begin{environ}{eec}
    $e_{1}$ & $\cdots$ & $e_{n}$
  \end{environ}}
  $\kern0.5pt, consult the types of the~$e_{i}$ to select the most efficient closure representation
  (recall our discussion in~\cref{sec:closure-representation}).
\end{compactitem}

\begin{wrapfigure}{l}[14pt]{0pt}
  \centering\small\hskip-10pt
  \begin{tabular}{@{}>{\ttfamily}lr@{}}
    \toprule
    \textnormal{\textbf{Query/Function}} &
    \multicolumn{1}{c@{}}{\textbf{Simplified}}
    \\
    \cmidrule(r){1-1}\cmidrule(l){2-2}
     \sql{SELECT\,\tpchcol{o}{orderkey,}}$\cdots$                              & 36\,010 \\
     \textSFii\textSFx\sql{\tpchcol{dispatch}{1}()}                            & 31\,851 \\
     ~~\textSFviii\textSFx\sql{$\ell_{1}$()}                                   & 18\,023 \\
     ~~\textSFxi~\textSFii\textSFx\sql{SELECT\,GREATEST($\cdots$}              &  4\,770 \\
     ~~\textSFviii\textSFx\sql{$\ell_{2}$()}                                   &  2\,923 \\
     ~~\textSFii\textSFx\sql{$\ell_{3}$()}                                     &     154 \\
    \bottomrule
  \end{tabular}
\end{wrapfigure}
\noindent
For the \PLSQL{} program of~\cref{fig:sql-f-o-PLSQL}, these simplifications
lead to the removal of~\sql{\tpchcol{dispatch}{2}()}
since the functional argument~\sql{comp} is statically known to be~\sql{GREATEST} in the present
example. Execution time is reduced by an additional~$11\,\%$ (see column~\textbf{Simplified}
above). We mention that the execution time now is
within~$19\,\%$ of a first-order formulation of the program---this
first-order variant is less flexible as it replaces the join with
(re-)configurable function table~\sql{COMPLETION} by an explicit
hard-wired~\sql{CASE} statement, however.

%
%



\medskip\noindent
\textbf{Avoiding Closure Construction.}
A closer look at the ``native'' row
of~\cref{tab:xquery-dyn-call-closures} shows that a growing number
of free variables only has moderate impact on BaseX' and Saxon's
native implementations of dynamic function calls: in the second-round
experiments, both processors expand the definitions of free variables inside
the called function's body, effectively avoiding the need for an environment.
\emph{Unfolding optimizations} of this kind can also benefit defunctionalization.

The core of such an inlining optimizer is a source-level query rewrite in which
closure construction and elimination cancel each other out:

{\small
$$
  \begin{array}{l}
    \meta{case}\
    \closure{$\ell_{}$}{
      \begin{environ}{eec}
        $e_{1}$ & ${\cdots}$ & $e_{n}$
      \end{environ}
    }\
    \meta{of}
    \\[-1ex]
    \qquad \vdots \\
    \quad \closure{$\ell_{}$}{
      \begin{environ}{eec}
        $\vardollar{}v_{1}$ & ${\cdots}$ & $\vardollar{}v_{n}$
      \end{environ}
    } \Rightarrow e
    \\[-0.5ex]
    \smash[b]{\qquad\vdots}
  \end{array}
  \quad \rightsquigarrow \quad
  \begin{array}{l@{~}c@{~}c@{~}l}
    \xquery{let} & \vardollar{}v_{1} & \xquery{:=} & e_{1} \\[-1ex]
                 & \vdots \\[-0.5ex]
                 & \vardollar{}v_{n} & \xquery{:=} & e_{n} \\
    \multicolumn{3}{l}{\xquery{return}~e}
  \end{array}
$$}

\noindent
As this simplification depends on the closure label~$\ell$ and the
environment contents~$e_{1}, \dots, e_{n}$ to be statically known at
the $\meta{case}\cdots\meta{of}$ site, the rewrite works in tandem with
unfolding transformations:
\begin{compactitem}
\item Replace \xquery{let}-bound variables by their definitions if the latter are
  considered simple (\emph{e.g.}, literals or closures with simple environment contents).
\item Replace applications of function literals or calls to user-defined non-recursive functions
  by the callee's body in which function arguments are \xquery{let}-bound.
\end{compactitem}
Defunctionalization and subsequent unfolding optimization transform
the \XQuery{} \xquery{group-by} example of~\cref{fig:xquery-h-o-opt-group-by} into
the first-order query of~\cref{fig:xquery-f-o-opt-group-by}.
In the optimized query, the dispatchers~\xquery{dispatch\_0}
and~\xquery{dispatch\_1} (cf.\ \cref{fig:xquery-f-o-group-by}) have been inlined.
The construction and elimination of closures with label~$\ell_{2}$ canceled
each other out.

\begin{figure}
  \centering\small
  \begin{lstlisting}[language=xquery]
let $fib := (0,1,1,2,3,5,8,13,21,34)
let $keys := for $x in $fib return $x mod 2
for $x in for $k in distinct-values($keys)
          let $group := $fib[((.) mod 2) = $k]
          return %\closure{$\ell_{1}$}{\begin{environ}{c} \$group \end{environ}}%
return element group { %\meta{case}% $x %\meta{of}%
                          %$\closure{$\ell_{1}$}{\begin{environ}{c} \$group \end{environ}} \Rightarrow$% $group
                     }
  \end{lstlisting}
  \caption{First-order \XQuery{} code for the example of~\cref{fig:xquery-h-o-opt-group-by}
    (defunctionalization and unfolding rewrite applied).}
  \label{fig:xquery-f-o-opt-group-by}
\end{figure}

Finally, the above mentioned simplifications succeed in removing the remaining
closures labeled~$\ell_{1}$, leaving us with closure-less code.
\cref{tab:inlining-opt} compares evaluation times for the original
defunctionalized \xquery{group-by} code and its optimized variants---all three
\XQuery{}~1.0 processors clearly benefit.

\begin{table}[t]
  \centering\small
  \begin{tabular}{@{}rccc@{}}
    \toprule
    & \textbf{Oracle} & \textbf{Berkeley\,DB} & \textbf{Sedna}
    \\
    \cmidrule(lr){2-2}\cmidrule(lr){3-3}\cmidrule(l){4-4}
    defunctionalization & 5.03 & 20.60           & 2.56 \\
    $+$       unfolding & 4.99 & \phantom{0}9.29 & 1.31 \\
    $+$ simplifications & 1.28 & \phantom{0}7.45 & 0.98 \\
    \bottomrule
  \end{tabular}
  \vskip5pt
  \caption{Impact of unfolding and simplifications on the evaluation of
    \xquery{group-by(\$seq,\,function(\$x)\,\mtt{\{}\,\$x\,mod\,100}\,\mtt{\})} for
    $\lvert\xquery{\$seq}\rvert = $ 10$^\mathbf{4}$.  Averaged wall-clock time measured in seconds.}
  \label{tab:inlining-opt}
\end{table}

\section{More Related Work}
\label{sec:related-work}

Query defunctionalization as described here builds on a body of work on
the removal of higher-order functions in programs written in functional
programming languages.  The representation of closures in terms of first-order
records has been coined as \emph{closure-passing style}~\cite{ClosureCodePointer}.
Dispatchers may be understood as mini-interpreters that inspect closures to
select the next program step (here: surrogate function) to execute, a
perspective due to Reynolds~\cite{DefuncReynolds}.  Our particular formulation
of defunctionalization relates to Tolmach and Oliva and their
work on translating ML to Ada~\cite{MLtoAda} (like the target query languages
we consider, Ada\,83 lacks code pointers).

The use of higher-order functions in programs can be normalized away if
specific restrictions are obeyed. Cooper~\cite{Cooper} studied such a
translation that derives \SQL{} queries from programs that have a flat
list (\emph{i.e.}, tabular) result type---this constraint rules out
tables of functions, in particular. Program normalization is a
runtime activity, however, that is not readily integrated with existing query engine
infrastructure.

With HOMES~\cite{HOMES}, Benedikt and Vu have developed higher-order
extensions to relational algebra and Core~\XQuery{} that add
abstraction (admitting queries of function type that accept queries
as parameters) as well as dynamic function calls (applying queries
to queries). HOMES' query processor alternates between regular
database-supported execution of query blocks inside Postgre\SQL{} or
BaseX and graph-based $\beta$-reduction outside a database system. In
contrast, defunctionalized queries may be executed while staying within
the context of the database kernel.

From the start, the design of FQL~\cite{FQL} relied on functions as
the primary query building blocks: following Backus' FP language, FQL
offers functional forms to construct new queries
out of existing functions. Buneman \emph{et al.}\ describe a general
implementation technique that evaluates FQL queries lazily. The central
notion is that of \emph{suspensions}, pairs~$\left<f, x\right>$
that represent the yet unevaluated application of function~$f$ to
argument~$x$. Note how~\xquery{group-by} in~\cref{fig:xquery-f-o-group-by}
mimics suspension semantics by returning closures (with label~$\ell_{1}$)
that only get evaluated (via~\xquery{dispatch\_0}) once a group's members
are required.

A tabular data model that permits function-valued columns has been
explored by Stonebraker \emph{et al.}~\cite{QUEL}. Such columns hold
QUEL expressions, represented either as query text or compiled plans.
Variables may range over QUEL values and an \sql{exec($e$)}~primitive is
available that spawns a separate query processor instance to evaluate
the QUEL-valued argument~$e$ at runtime.

Finally, the Map-Reduce model~\cite{MapReduce} for massively
distributed query execution successfully adopts a function-centric
style of query formulation. Functions are not first-class, though:
first-order user-defined code is supplied as arguments to two built-in
functions~$\var{map}$ and~$\var{reduce}$---Map-Reduce builds on
higher-order function constants but lacks function variables.

Defunctionalized \XQuery{} queries that rely on an element-based
representation of closures create \XML{} fragments (closure
construction) whose contents are later extracted via \xquery{child}
axis steps (closure elimination). When node construction and
traversal meet like this, the creation of intermediate fragments can
be avoided altogether. Such fusion techniques have been specifically
described for \XQuery{}~\cite{XQFusion}. Fusion, jointly with function
inlining as proposed in~\cite{XQFunctionInlining}, thus can implement
the $\meta{case}\cdots\meta{of}$ cancellation optimization discussed
in~\cref{sec:experiments}.  If cancellation is not possible, \XQuery{}
processors can still benefit from the fact that node identity and document
order are immaterial in the remaining intermediate fragments~\cite{eXrQuy}.

%
%
%
\section{Closure}
\label{sec:conclusions}

We argue that a repertoire of literal function values, higher-order
functions, and functions in data structures can lead to particularly
concise and elegant formulations of queries. Query defunctionalization
enables off-the-shelf first-order database engines to support such a
function-centric style of querying. Cast in the form of a syntax-directed
transformation of queries, defunctionalization is non-invasive and affects
the query processor's front-end only (a simple preprocessor will also yield
a workable implementation).  Experiments show that the technique does
not introduce an undue runtime overhead.

Query defunctionalization applies to any query language that
\begin{inparaenum}[(1)]
\item offers aggregate data structures suitable to represent closures
  and
\item implements case discrimination based on the contents of such
  aggregates.
\end{inparaenum}
These are light requirements met by many languages beyond \XQuery{} and
\PLSQL{}. It is hoped that our discussion of query defunctionalization
is sufficiently self-contained such that it can be carried over to
other languages and systems.

\medskip\noindent
\textbf{Acknowledgment.} We dedicate this work to the memory of
John C.\ Reynolds (\Cross\,April~2013).

\balance
\bibliographystyle{abbrv}

\begin{thebibliography}{10}
  \bibitem{BerkeleyDBXML}
  {Oracle} {Berkeley} {DB} {XML}.
  \newblock
    \url{http://www.oracle.com/technetwork/products/berkeleydb/index-083851.html}.

  \bibitem{PostgreSQL}
  {PostgreSQL}~9.2.
  \newblock \url{http://www.postgresql.org/docs/9.2/}.

  \bibitem{Saxon}
  {Saxon}.
  \newblock \url{http://saxon.sourceforge.net/}.

  \bibitem{OraclePLSQL}
  {\em Oracle Database PL/SQL Language Reference---11g Release~1~(11.1)}, 2009.

  \bibitem{ClosureCodePointer}
  A.~Appel and T.~Jim.
  \newblock {Continuation-Passing, Closure-Passing Style}.
  \newblock In {\em Proc.\ POPL}, 1989.

  \bibitem{IntroFunctionalProgramming}
  R.~Bird and P.~Wadler.
  \newblock {\em {Introduction to Functional Programming}}.
  \newblock Prentice Hall, 1988.

  \bibitem{Thunking}
  A.~Bloss, P.~Hudak, and J.~Young.
  \newblock {Code Optimizations for Lazy Evaluation}.
  \newblock {\em Lisp and Symbolic Computation}, 1(2), 1988.

  \bibitem{XQuery1.0}
  S.~Boag, D.~Chamberlin, M.~Fern\'{a}ndez, D.~Florescu, J.~Robie, and
    J.~Sim\'{e}on.
  \newblock {XQuery}~1.0: {An} {XML} {Query} {Language}.
  \newblock {W3C} {Recommendation}, 2010.

  \bibitem{MonetDBXQuery}
  P.~Boncz, T.~Grust, M.~van Keulen, S.~Manegold, J.~Rittinger, and J.~Teubner.
  \newblock {MonetDB/XQuery: A Fast XQuery Processor Powered by a Relational
    Engine}.
  \newblock In {\em Proc.\ SIGMOD}, 2006.

  \bibitem{FQL}
  P.~Buneman, R.~Frankel, and R.~Nikhil.
  \newblock {An} {Implementation} {Technique} for {Database} {Query} {Languages}.
  \newblock {\em ACM TODS}, 7(2), 1982.

  \bibitem{InitialXQuery1.0}
  D.~Chamberlin, D.~Florescu, J.~Robie, J.~Sim\'{e}on, and M.~Stefanescu.
  \newblock {XQuery:} {A} {Query} {Language} for {XML}.
  \newblock {W3C} {Working} {Draft}, 2001.

  \bibitem{Cooper}
  E.~Cooper.
  \newblock {The Script-Writers Dream: How to Write Great SQL in Your Own
    Language, and be Sure it Will Succeed.}
  \newblock In {\em Proc.\ DBPL}, 2009.

  \bibitem{MapReduce}
  J.~Dean and S.~Ghemawat.
  \newblock {MapReduce}: {Simplified} {Data} {Processing} on {Large} {Clusters}.
  \newblock In {\em Proc.\ OSDI}, 2004.

  \bibitem{EdgeApproach}
  D.~Florescu and D.~Kossmann.
  \newblock {Storing} and {Querying} {XML} {Data} {Using} an {RDBMS}.
  \newblock {\em IEEE Data Engineering Bulletin}, 22(3), 1999.

  \bibitem{Sedna}
  A.~Fomichev, M.~Grinev, and S.~Kuznetsov.
  \newblock Sedna: A {Native} {XML} {DBMS}.
  \newblock In {\em Proc.\ SOFSEM}, 2006.

  \bibitem{XQFunctionInlining}
  M.~Grinev and D.~Lizorkin.
  \newblock {XQuery} {Function} {Inlining} for {Optimizing} {XQuery} {Queries}.
  \newblock In {\em Proc.\ ADBIS}, 2004.

  \bibitem{BaseX}
  C.~Gr{\"u}n, A.~Holupirek, and M.~Scholl.
  \newblock {Visually} {Exploring} and {Querying} {XML} with {BaseX}.
  \newblock In {\em Proc.\ BTW}, 2007.
  \newblock \url{http://basex.org}.

  \bibitem{VersatileComprehensions}
  T.~Grust.
  \newblock {Monad Comprehensions: A Versatile Representation for Queries}.
  \newblock In {\em {The Functional Approach to Data Management -- Modeling,
    Analyzing and Integrating Heterogeneous Data}}. Springer, 2003.

  \bibitem{eXrQuy}
  T.~Grust, J.~Rittinger, and J.~Teubner.
  \newblock {eXrQuy:} {Order} {Indifference} in {XQuery}.
  \newblock In {\em Proc.\ ICDE}, 2007.

  \bibitem{FunctionsAreDataToo}
  T.~Grust, N.~Schweinsberg, and A.~Ulrich.
  \newblock {Functions} are {Data} {Too} ({Software} {Demonstration}).
  \newblock In {\em Proc.\ VLDB}, 2013.

  \bibitem{LambdaLifting}
  T.~Johnsson.
  \newblock Lambda {Lifting}: {Transforming} {Programs} to {Recursive}
    {Equations}.
  \newblock In {\em Proc.\ IFIP}, 1985.

  \bibitem{XQFusion}
  H.~Kato, S.~Hidaka, Z.~Hu, K.~Nakano, and I.~Yasunori.
  \newblock {Context}-{Preserving} {XQuery} {Fusion}.
  \newblock In {\em Proc.\ APLAS}, 2010.

  \bibitem{ClosureOriginal}
  P.~Landin.
  \newblock {The Mechanical Evaluation of Expressions}.
  \newblock {\em The Computer Journal}, 6(4):308--320, 1964.

  \bibitem{OracleXQuery}
  Z.~Liu, M.~Krishnaprasad, and A.~V.
  \newblock Native {XQuery} {Processing} in {Oracle} {XMLDB}.
  \newblock In {\em Proc.\ SIGMOD}, 2005.

  \bibitem{Dremel}
  S.~Melnik, A.~Gubarev, J.~Long, G.~Romer, S.~Shivakumar, M.~Tolton, and
    T.~Vassilakis.
  \newblock {Dremel}: {Interactive} {Analysis} of {Web-Scale} {Datasets}.
  \newblock {\em PVLDB}, 3(1), 2010.

  \bibitem{DefuncPolymorphic}
  F.~Pottier and N.~Gauthier.
  \newblock {Polymorphic} {Typed} {Defunctionalization}.
  \newblock In {\em Proc.\ POPL}, 2004.

  \bibitem{DefuncReynolds}
  J.~Reynolds.
  \newblock {Definitional} {Interpreters} for {Higher-Order} {Programming}
    {Languages}.
  \newblock In {\em Proc.\ ACM}, 1972.

  \bibitem{XQuery3.0}
  J.~Robie, D.~Chamberlin, J.~Sim\'{e}on, and J.~Snelson.
  \newblock {XQuery}~3.0: {An} {XML} {Query} {Language}.
  \newblock {W3C} {Candidate} {Recommendation}, 2013.

  \bibitem{LinkedClosures}
  Z.~Shao and A.~Appel.
  \newblock {Space-Efficient Closure Representations}.
  \newblock In {\em Proc.\ Lisp and Functional Programming}, 1994.

  \bibitem{SpaceLeaks}
  Z.~Shao and A.~Appel.
  \newblock {Efficient and Safe-for-Space Closure Conversion}.
  \newblock {\em ACM TOPLAS}, 22(1), 2000.

  \bibitem{SQL:1999}
  {\em {Database} {Language} {SQL}---{Part 2}: {Foundation} {(SQL/Foundation)}}.
  \newblock ANSI/ISO/IEC~9075, 1999.

  \bibitem{QUEL}
  M.~Stonebraker, E.~Anderson, E.~Hanson, and B.~Rubenstein.
  \newblock {QUEL} as a {Data} {Type}.
  \newblock In {\em Proc.\ SIGMOD}, 1984.

  \bibitem{MLtoAda}
  A.~Tolmach and D.~Oliva.
  \newblock {From} {ML} to {Ada:} {Strongly}-{Typed} {Language}
    {Interoperability} via {Source} {Translation}.
  \newblock {\em J. Funct. Programming}, 8(4), 1998.

  \bibitem{TPC-H}
  Transaction Processing Performance Council.
  \newblock {\em {TPC-H}, a {Decision-Support} {Benchmark}}.
  \newblock \url{http://tpc.org/tpch/}.

  \bibitem{HOMES}
  H.~Vu and M.~Benedikt.
  \newblock {HOMES}: {A} {Higher-Order} {Mapping} {Evalution} {System}.
  \newblock {\em PVLDB}, 4(12), 2011.

  \bibitem{XRPC}
  Y.~Zhang and P.~Boncz.
  \newblock {XRPC:} {Interoperable} and {Efficient} {Distributed} {XQuery}.
  \newblock In {\em Proc.\ VLDB}, 2007.

  \end{thebibliography}
\begin{spacing}{0.5}

\end{spacing}


\clearpage
\appendix
\section*{Appendix}

\section{Defunctionalization for \XQuery{}}
\label{sec:defunctionalization-for-xquery}

\begin{figure*}
  \centering\small
$
\begin{array}{rcl@{}c@{}}
  \decl\llbracket|declare function|~\var{name}|($|\var{x_1}|,|\dots|,$|\var{x_n}|)|~|{|~e~|}|\rrbracket
  & = &
  |declare function|~\var{name}|($|x_{1}|,|\dots|,$|x_{n}|)|~|{|~\expr\llbracket e\rrbracket~|}|
  \\[5ex]
  \expr\llbracket|for|~\vardollar{}v~|in|~e_{1}~|return|~e_{2}\rrbracket
  & = &
  |for|~\vardollar{}v~|in|~\expr\llbracket e_{1}\rrbracket~|return|~\expr\llbracket e_{2}\rrbracket
  \\
  \expr\llbracket|let|~\vardollar{}v~|:=|~e_{1}~|return|~e_{2}\rrbracket
  & = &
  |let|~\vardollar{}v~|:=|~\expr\llbracket e_{1}\rrbracket~|return|~\expr\llbracket e_{2}\rrbracket
  \\
  \expr\llbracket\vardollar{}v\rrbracket
  & = &
  \vardollar{}v
  \\
  \expr\llbracket\xquery{if~(}e_{1}\xquery{)~then}~e_{2}~\xquery{else}~e_3\rrbracket
  & = &
  |if (|\expr\llbracket e_{1}\rrbracket|) then|~\expr\llbracket e_{2}\rrbracket~|else|~\expr\llbracket e_{3}\rrbracket
  \\
  \expr\llbracket|(|e_{1}|,|\dots|,|e_{n}|)|\rrbracket
  & = &
  |(|\expr\llbracket e_{1}\rrbracket|,|\dots|,|\expr\llbracket e_{n}\rrbracket|)|
  \\
  \expr\llbracket e|/|a|::|t\rrbracket
  & = &
  \expr\llbracket e\rrbracket|/|a|::|t
  \\
  \expr\llbracket|element|~n~|{|~e~|}|\rrbracket
  & = &
  |element|~n~|{|~\expr\llbracket e\rrbracket~|}|
  \\
  \expr\llbracket e_1 |[| e_2 |]| \rrbracket
  & = &
  \expr\llbracket e_1 \rrbracket |[| \expr\llbracket e_2 \rrbracket |]|
  \\
  \expr\llbracket |.| \rrbracket
  & = &
  |.|
  \\
  \expr\llbracket\var{name}|(|e_1|,|\dots|,| e_n|)|\rrbracket
  & = &
  \var{name}|(|\expr\llbracket e_{1}\rrbracket|,|\dots|,|\expr\llbracket e_{n}\rrbracket|)|
  \\
  \labelcase{case:literal}
  \expr\llbracket|function(|\vardollar{}x_1\,|as|\,t_{1}|,|\dots|,|\vardollar{}x_n\,|as|\,t_{n}|)|\,|as|\,t\,|{|~e~|}|\rrbracket
  & = &
  \closure{$\ell_{}$}{
    \begin{environ}{eec}
      \vardollar{}$v_{1}$ & $\cdots$ & \vardollar{}$v_{m}$
    \end{environ}
  }
  &
  \labelcref{case:literal}
  \\
  & &
  \begin{array}{@{}l@{}}
    \begin{array}{@{}r@{~}c@{~}l@{}}
      \Dispatch{n} & \shortleftarrow & \Dispatch{n} \cup \left\{ \var{branch} \right\} \\
      \var{Lifted} & \shortleftarrow & \var{Lifted} \cup \left\{ \var{lifted} \right\}
    \end{array} \\
    \textbf{where}
    \\
    \begin{array}[t]{@{~}r@{~}c@{~}l}
       \ell                      & = & \var{label}(n) \\
       \vardollar{}v_{1},\dots,\vardollar{}v_{m}  & = & \var{fv}(|function(|\vardollar{}x_1\,|as|\,t_{1}|,|\dots|,|\vardollar{}x_n\,|as|\,t_{n}|)|\,|as|\,t\,|{|~e~|}|) \\
       \var{branch}              & = & \closure{$\ell_{}$}{
                                         \begin{environ}{eec}
                                             \vardollar{}$v_{1}$ & $\cdots$  & \vardollar{}$v_{m}$
                                         \end{environ}} \Rightarrow \\
                                 &   & \quad \ell|(|\vardollar{}|b|_1|,| \dots|,| \vardollar{}|b|_n|,| \vardollar{}v_1|,| \dots|,| \vardollar{}v_m|)| \\
       \var{lifted}              & = & |declare function|~\ell|(| \\
                                 &   & \quad\vardollar{}x_1\,|as|\,t_{1}|,| \dots|,| \vardollar{}x_n\,|as|\,t_{n}|,| \vardollar{}v_1|,| \dots|,| \vardollar{}v_m|)|\,|as|\,t \\
                                 &   & |{|~\expr\llbracket e \rrbracket~|};| \\
    \end{array}
  \end{array}
  \\[5ex]
  \labelcase{case:name}
  \expr\llbracket name|#|n\rrbracket
  & = &
  \closure{$\ell_{}$}{\emptyenv} 
  &
  \labelcref{case:name}
  \\
  & &
  \begin{array}{@{}l@{}}
    \Dispatch{n} \shortleftarrow \Dispatch{n} \cup \left\{ \var{branch} \right\} \\
    \textbf{where} \\
    \begin{array}[t]{@{~}r@{~}c@{~}l}
      \ell           & = & \var{label}(n) \\
      \var{branch}   & = & \closure{$\ell$}{\emptyenv} \Rightarrow \var{name}|(|\vardollar{}|b|_{1}|,| \dots|,| \vardollar{}|b|_{n}|)| \\
    \end{array}
  \end{array}
  \\[5ex]
  \labelcase{case:apply}
  \expr\llbracket e|(|e_1|,|\dots|,| e_n|)|\rrbracket
  & = &
  |dispatch_|{n}|(|\expr\llbracket e\rrbracket|,|\expr\llbracket e_{1}\rrbracket|,|\dots|,|\expr\llbracket e_{n}\rrbracket|)|
  &
  \labelcref{case:apply}
  \\[5ex]
  \query\llbracket d_1|;|\dots|;|\var{d_n}|;|\,e\rrbracket
  & = &
  \begin{array}[t]{@{}l}
    \forall\,i \in \var{dom}(\var{Dispatch}){:}~\var{declare\_dispatch}(i, \Dispatch{i}) \\
    \var{Lifted} \\
    \decl\llbracket d_1\rrbracket|;|\dots|;|
    \decl\llbracket d_n\rrbracket|;|\,
    \expr\llbracket e\rrbracket
  \end{array}
\end{array}
$
  \caption{Defunctionalization of \XQuery{}~3.0 function declarations ($\decl$),
    expressions ($\expr$) and queries ($\query$).}
  \label{fig:defun-rewrite-comprehension}
\end{figure*}

This appendix elaborates the details of defunctionalization
for~\XQuery{}~3.0.  The particular formulation we follow here is a
deliberate adaptation of the transformation as it has been described by
Tolmach and Oliva~\cite{MLtoAda}.

We specify defunctionalization in terms of a syntax-directed traversal,
$\query\llbracket e\rrbracket$, over a given \XQuery{}~3.0 source query~$e$
(conforming to~\cref{fig:source-grammar}). In general, $e$ will contain a series of
function declarations which precede one main expression to evaluate. $\query$
calls on the auxiliary~$\decl\llbracket\cdot\rrbracket$
and~$\expr\llbracket\cdot\rrbracket$ traversals to jointly transform
declarations and expressions---this makes~$\query$ a
\emph{whole-query transformation}~\cite{DefuncPolymorphic} that needs to
see the input query in its entirety. All three traversal schemes are defined
in~\cref{fig:defun-rewrite-comprehension}.

$\expr$ features distinct cases for each of the syntactic constructs
in the considered \XQuery{}~3.0 subset. However, all cases but those
labeled~\labelcref{case:literal}--\labelcref{case:apply} merely
invoke the recursive traversal of subexpressions, leaving their input
expression intact otherwise.  The three cases implement the transformation
of literal functions, named function references, and dynamic function
calls.  We will now discuss each of them in turn.

\medskip\noindent
\textbf{\Cref{case:literal}:~Literal Functions.}
Any occurrence of a literal function, say~$f =
\xquery{function(\$$x_{1}$,$\dots$,\$$x_{n}$)\,\{\,$e$\,\}}$, is
replaced by a closure constructor. Meta-level function~$\var{label}()$
generates a unique label~$\ell$ which closure elimination will later use to
identify~$f$ and evaluate its body expression~$e$; see~\cref{case:apply}
below.
The evaluation of~$e$ depends on its free variables, \emph{i.e.}, those
variables that have been declared in the lexical scope enclosing~$f$. We use
meta-level function~$\var{fv}()$ to identify these
variables~$\xquery{\$$v_{1}$}, \dots, \xquery{\$$v_{m}$}$ and save their
values in the closure's environment.  At runtime, when the closure constructor
is encountered in place of~$f$, the closure thus captures the state required
to properly evaluate subsequent applications of~$f$
(recall~\cref{sec:closures}).  Note that defunctionalization does not rely on
functions to be pure: side-effects caused by body~$e$ will also be
induced by~$\expr\llbracket e\rrbracket$.

To illustrate, consider the following \XQuery~3.0 snippet, taken from
the \xquery{group-by} example in~\cref{fig:xquery-h-o-group-by}:
\begin{center}
\begin{minipage}{5.6cm}
\begin{lstlisting}[numbers=none]
for $k in distinct-values($keys)
return
  function() { $seq[$key(.) = $k] } %\enskip\textnormal{.}%
\end{lstlisting}
\end{minipage}
\end{center}
We
have~$\var{fv}(\xquery{function()\,\{\,\$seq[\$key(.)\,=\,\$k]\,\}}) =
\xquery{\$k}$, $\xquery{\$key}$, $\xquery{\$seq}$. According to~$\expr$
and~\cref{case:literal} in particular, the snippet thus defunctionalizes to
\begin{center}
\begin{minipage}{5.5cm}
\begin{lstlisting}[numbers=none]
for $k in distinct-values($keys)
return
    %\closure{$\ell_1$}{
      \begin{environ}{eec}
        \xquery{\$k} & \xquery{\$key} & \xquery{\$seq}
      \end{environ}
    }
\end{lstlisting}
\end{minipage}
\end{center}
where~$\ell_{1}$ denotes an arbitrary yet unique label.

If we assume that the free variables are defined as in the example
of~\cref{fig:xquery-h-o-group-by}, the defunctionalized variant of the snippet
will evaluate to a sequence of two closures:
$$
\xquery{(}\closure{$\ell_{1}$}{
  \begin{environ}{eec}
    \xquery{0} & \closure{$\ell_{2}$}{\emptyenv} & \xquery{(0,1,1,2,$\dots$)}
  \end{environ}
}\xquery{, } \closure{$\ell_{1}$}{
  \begin{environ}{eec}
    \xquery{1} & \closure{$\ell_{2}$}{\emptyenv} & \xquery{(0,1,1,2,$\dots$)}
  \end{environ}
}\xquery{)} \enskip.
$$
These closures capture the varying values~\xquery{0}, \xquery{1} of
the free iteration variable~\xquery{\$k} as well as the invariant
values of~\xquery{\$key} (bound to a function and thus represented
in terms of a closure with label~$\ell_{2}$) and~\xquery{\$seq} ($=
\xquery{(0,1,1,2,$\dots$)}$).

Since we will use label~$\ell_{1}$ to identify the body of the
function literal \xquery{function()\,\{\,\$seq[\$key(.)\,=\,\$k]\,\}},
\cref{case:literal} saves this label/body association in terms of
a~$\meta{case}\cdots\meta{of}$ branch (see the assignment to~$\var{branch}$
in~\cref{fig:defun-rewrite-comprehension}).  We will shed more light
on~$\var{branch}$ and~$\var{lifted}$ when we discuss~\cref{case:apply} below.

\medskip\noindent
\textbf{\Cref{case:name}: Named Function References.}  Any occurrence of an
expression~\xquery{$\var{name}$\#$n$}, referencing function~$\var{name}$
of arity~$n$, is replaced by a closure constructor with a unique
label~$\ell$. In~\XQuery, named functions are \emph{closed} as they are
exclusively declared in a query's top-level scope---either in the query prolog
or in an imported module~\cite{XQuery3.0}---and do not contain free variables.
In~\cref{case:name}, the constructed closures thus have empty environments.  As
before, a~$\meta{case}\cdots\meta{of}$ branch is saved that associates
label~$\ell$ with function~$\var{name}$.

\medskip\noindent
\textbf{\Cref{case:apply}: Dynamic Function Calls.}
In case of a dynamic function call~\xquery{$e$($e_{1}$,$\dots$,$e_{n}$)}, we
know that expression~$e$ evaluates to \emph{some} functional value
(otherwise~$e$ may not occur in the role of a function and be applied
to arguments).\footnote{Note that $\expr\llbracket\cdot\rrbracket$
defines a separate case for static function calls of the
form~\xquery{$\var{name}$($e_{1}$,$\dots$,$e_{n}$)}.} Given our
discussion of~\cref{case:literal,case:name}, in a defunctionalized
query, $e$ will thus evaluate to a closure, say
\closure{$\ell$}{
  \begin{environ}{eec}
    $x_{1}$ & $\cdots$ & $x_{m}$
  \end{environ}
} ($m \geqslant 0$), that represents some function~$f$.

In the absence of code pointers, we delegate the invocation of the function
associated with label~$\ell$ to a \emph{dispatcher}, an auxiliary routine
that defunctionalization adds to the prolog of the transformed query.
The dispatcher
\begin{compactenum}[(\itshape i\/\upshape)]
\item receives the closure as well as $e_{1}, \dots, e_{n}$ (the arguments
  of the dynamic call) as arguments, and then
  \label{enum:dispatch1}
\item uses $\meta{case}\cdots\meta{of}$ to select the branch associated
  with label~$\ell$.
  \label{enum:dispatch2}
\item The branch unpacks the closure environment to extract the bindings of the $m$~free
  variables (if any) that were in place when~$f$ was defined, and finally
  \label{enum:dispatch3}
\item invokes a surrogate function that contains the body of the original
  function~$f$, passing the $e_{1}, \dots, e_{n}$ along with the extracted
  bindings (the surrogate function thus has arity $n + m$).
  \label{enum:dispatch4}
\end{compactenum}

\begin{figure}
  \centering\small
  \begin{tabular}{@{}l@{}}
    $\var{declare\_dispatch}(n, \{\var{case}_{1}, \dots, \var{case}_{k}\}) \equiv$
    \\
    \hspace{5pt} 
    \begin{minipage}{0.95\linewidth}
    \begin{lstlisting}
declare function dispatch_%$n$%(
  $clos as %\meta{closure}%,
  $b%$_{1}$% as item()*,%$\dots$%, $b%$_{n}$% as item()*) as item()*
{
  %\meta{case}% $clos %\meta{of}%
    %$\var{case}_{1}$%
      %\raisebox{-2pt}{\smash{$\vdots$}}%
    %$\var{case}_{k}$%
};
    \end{lstlisting}
    \end{minipage}
  \end{tabular}
  \caption{Declaring a dispatcher for $n$-ary functional values.}
  \label{fig:declare-dispatch}
\end{figure}

\medskip\noindent
\emph{Re~(\labelcref{enum:dispatch1}) and~(\labelcref{enum:dispatch2}).} In our
formulation of defunctionalization for \XQuery{}, a dedi\-cated dispatcher
is declared for all literal functions and named function references that are
of the same arity.  The~$\meta{case}\cdots\meta{of}$ branches for the dispatcher
for arity~$n$ are collected in set $\Dispatch{n}$ while~$\expr$ traverses the input
query (\cref{case:literal,case:name} in~\cref{fig:defun-rewrite-comprehension}
add a branch to $\Dispatch{n}$ when an $n$-ary functional value is transformed).
Once the traversal is complete, $\query$ adds the dispatcher routine to the
prolog of the defunctionalized query through~$\var{declare\_dispatch(n, \Dispatch{n})}$.
This meta-level function, defined in~\cref{fig:declare-dispatch}, emits
the routine~\xquery{dispatch\_$n$} which receives closure~\xquery{\$clos}
along with the $n$~arguments of the original dynamic call. Discrimination on the
label~$\ell$ stored in~\xquery{\$clos} selects the associated branch.
Because~\xquery{dispatch\_$n$}
dispatches calls to \emph{any} $n$-ary function in the original query, we declare
it with a polymorphic signature featuring \XQuery{}'s most polymorphic type~\xquery{item()*}.
The \PLSQL{} variant of defunctionalization, discussed in \cref{sec:defunctionalization-plsql},
relies on an alternative approach that uses typed dispatchers.

\medskip\noindent
Any occurrence of a dynamic function call~\xquery{$e$($e_{1}$,$\dots$,$e_{n}$)}
is replaced by a static call to the appropriate dispatcher
\xquery{dispatch\_$n$}.

\cref{fig:xquery-f-o-group-by} (in the main text) shows the defunctionalized
query for the \XQuery{}~\xquery{group-by} example
of~\cref{fig:xquery-h-o-group-by}. The original query contained literal
functions of arity~$0$ (in~\cref{line:capture}) as well as arity~$1$
(in~\cref{line:mod}). Following~\cref{case:literal}, both have been replaced
by closure constructors (with labels~$\ell_{1}$ and~$\ell_{2}$, respectively,
see~\cref{line:capture-closure,line:mod-closure}
in~\cref{fig:xquery-f-o-group-by}).
\xquery{function(\$x)\,}|{|\xquery{\,\$x\,mod\,2\,}|}| is closed: its closure
(label~$\ell_{2}$) thus contains an empty environment. Dynamic calls to both
functions have been replaced by static calls to the
dispatchers~\xquery{dispatch\_0} or~\xquery{dispatch\_1}.  For the present
example, $\Dispatch{0}$ and $\Dispatch{1}$ were singleton sets such that both
dispatchers contain~$\meta{case}\cdots\meta{of}$ expressions with one branch
only.  (For an example of a dispatcher with three branches, refer to the
\PLSQL{} function~\sql{dispatch\_1} in~\cref{fig:sql-f-o-PLSQL},
\cref{line:dispatch1}.)

\medskip\noindent
\emph{Re~(\labelcref{enum:dispatch3}) and~(\labelcref{enum:dispatch4}).}
Inside its dispatcher, the case branch for the closure
\closure{$\ell$}{
  \begin{environ}{eec}
    $x_{1}$ & $\cdots$ & $x_{m}$
  \end{environ}}
for function~$f$ invokes the associated surrogate function, also
named~$\ell$. The original arguments~$e_{1},\dots,e_{n}$ are passed
along with the $x_{1},\dots,x_{m}$. Surrogate function~$\ell$
incorporates $f$'s body expression and can thus act as a ``stand-in''
for~$f$. We declare the surrogate function with the same argument
and return types as~$f$---see the types $t$ and $t_{1},\dots,t_{n}$
in~\cref{case:apply} of~\cref{fig:defun-rewrite-comprehension}. The
specific signature for~$\ell$ ensures that the original semantics
of~$f$ are preserved (this relates to \XQuery{}'s \emph{function conversion
rules}~\cite[\S{}3.1.5.2]{XQuery3.0}).

While~$f$ contained $m$ free variables, $\ell$ is a closed function as it
receives the $m$ bindings as explicit additional function parameters
(surrogate function~$\ell$ is also known as the \emph{lambda-lifted} variant
of~$f$~\cite{LambdaLifting}). When \cref{case:literal} transforms a literal
function, we add its surrogate to the set $\var{Lifted}$ of function
declarations. When~\cref{case:name} transforms the named
reference~\xquery{$\var{name}$\#$n$}, $\var{Lifted}$ remains unchanged: the
closed function~$\var{name}$ acts as its own surrogate because there are no
additional bindings to pass. Again, once the traversal is complete, $\query$
adds the surrogate functions in set~$\var{Lifted}$ to the prolog of the
defunctionalized query. Returning to~\cref{fig:xquery-f-o-group-by}, we find
the two surrogate functions~$\ell_{1}$ and~$\ell_{2}$ at the top of the query
prolog (\crefrange{line:surrogate-start}{line:surrogate-end}).
\end{document}